\def\gtorder{\mathrel{\raise.3ex\hbox{$>$}\mkern-14mu
    \lower0.6ex\hbox{$\sim$}}}
\def\ltorder{\mathrel{\raise.3ex\hbox{$<$}\mkern-14mu
    \lower0.6ex\hbox{$\sim$}}}

\documentclass[fleqn,usenatbib]{mnras}

\usepackage[T1]{fontenc}

\DeclareRobustCommand{\VAN}[3]{#2}
\let\VANthebibliography\thebibliography
\def\thebibliography{\DeclareRobustCommand{\VAN}[3]{##3}\VANthebibliography}

\usepackage{graphicx}	% Including figure files
\usepackage{amsmath}	% Advanced maths commands
\usepackage{amssymb}	% Extra maths symbols
\usepackage{soul}
\usepackage{newtxtext,newtxmath}
\title[Galactic Morphologies at Cosmic Dawn]
  {Emergence of Galactic Morphologies at Cosmic Dawn: Input from Numerical Modeling}

\author[Bi, Shlosman and Romano-Diaz]{
Da Bi,$^{1}$\thanks{E-mail: dbi224@g.uky.edu}
Isaac Shlosman,$^{1,2}$
Emilio Romano-Díaz$^{3}$
\\
$^{1}$Department of Physics and Astronomy, University of Kentucky, Lexington, KY 40506-0055, USA\\
$^{2}$Theoretical Astrophysics, Graduate School of Science, Osaka University, Osaka, Japan\\
$^{3}$Argelander-Institute for Astronomy, Bonn, Germany
}

\date{Accepted XXX. Received YYY; in original form ZZZ}

\pubyear{2021}

\begin{document}
\label{firstpage}
\pagerange{\pageref{firstpage}--\pageref{lastpage}}
\maketitle

% Abstract of  the paper
\begin{abstract}
We employ a series of high-resolution zoom-in cosmological simulations to analyze the emerging morphology of main galaxies in dark matter halos at redshifts $z\gtorder 2$. We choose DM halos of {\it similar} masses of ${\rm log}\,M_{\rm vir}/{\rm M_\odot}\sim 11.65\pm 0.05$ at the target redshifts of $z_{\rm f} = 6$, 4 and 2. {  The rationale for this choice, among others, allows us to analyze how the different growth rate in these DM halos propagates down to galaxy scales and affects their basic parameters. Halos were embedded in high or low overdensity regions, and two different versions of a galactic wind feedback have been employed --- a strong one and energetically weak one. Our main results are: (1) Although our galaxies evolve in different epochs, their global parameters (e.g., baryonic masses and sizes) remain within a narrow range. Their morphology, kinematics and stellar populations differ substantially, yet all of them host sub-kpc stellar bars; (2) The star formation rates (SFRs) appear higher for larger $z_{\rm f}$, in tandem with their energy and momentum feedback; (3) The stellar kinematics allowed separation of bulge from the stellar spheroid. The existence of disk-like bulges has been revealed based on stellar surface density and photometry, but displayed a mixed disk-like and classical bulges based on their kinematics. The bulge-to-total mass ratios appear independent of the last merger time for all $z_{\rm f}$.  The stellar spheroid-to-total mass ratios of these galaxies lie in the range of $\sim 0.5-0.8$; (4) The synthetic redshifted, pixelized and PSF-degraded JWST images allow to detect stellar disks at all $z_{\rm f}$. Some bars disappear in degraded images, but others remain visible; (5) Based on the kinematic decomposition, for stellar disks separated from bulges and spheroids. we observe that rotational support in disks depends on the feedback type, but increases with decreasing $z_{\rm f}$; (6) Finally, the ALMA images detect disks at all $z_{\rm f}$, but their spiral structure is only detectable in $z_{\rm f}=2$ galaxies.} We also find that the galaxies follow the Tully-Fisher relation most of their evolution, being separated only by the galactic wind feedback.   
\end{abstract}

% Select between one and six entries from the list of approved keywords.
% Don't make up new ones.
\begin{keywords}
Methods: numerical -- galaxies: abundances --- galaxies: evolution --- galaxies: haloes --- galaxies:high-redshift -- galaxies: interactions
\end{keywords}

%%%%%%%%%%%%%%%%%%%%%%%%%%%%%%%%%%%%%%%%%%%%%%%%%%

%%%%%%%%%%%%%%%%% BODY OF PAPER %%%%%%%%%%%%%%%%%%

%%%%%%%%%%%%%%%%%%%%%%%%
\section{Introduction}
\label{sec:intro}
%%%%%%%%%%%%%%%%%%%%%%%%

Understanding galaxy evolution lies at the forefront of the contemporary astronomy research. Observations have continuously pushed the detected galaxy redshift limit, which now stands at $z\sim 11$ \citep[e.g.,][]{oesch16}, by applying mostly two methods --- searching for Ly\,$\alpha$ emitters (LAEs) and Lyman Break galaxies (LBGs) \citep[e.g.,][]{cass11,konno14,madau14,bouw15,hari18,ono18}. Most of the surveys of high-$z$ galaxies detect only the brightest galaxies \cite[e.g.,][]{forst06,genz08}, but using gravitational lensing and other techniques, fainter galaxies can be targeted \citep[e.g.,][]{kawa16,bouw17,lotz17,ishi18,shibu21}.  

{  Galactic morphology in the contemporary universe results from the  convergence of a long list of physical processes, not all of them yet understood and quantified. Among these are the role of environment and the parent DM halos, as well as the stellar and the AGN feedback, local and global instabilities in the stellar and gaseous disks, and origin of stellar bulges and bars.}

In this work we investigate the emergence of galactic morphologies at the Cosmic Dawn, at redshifts $z\gtorder 2$. Using very high-resolution zoom-in simulations, We focus on galaxy evolution within similar mass dark matter (DM) halos, log\,$M_{\rm vir}\sim 11.65\pm 0.05\,{\rm M_\odot}$, terminating their evolution at three target redshifts, $z_{\rm f} = 6$, 4 and 2. This evolution encompasses the reionization epoch, $z\gtorder 6$, and the subsequent time period of $\sim 2.5$\,Gyr, when the star formation in the universe peaks. Overall, evolution for $z\sim 9-2$ is analyzed. 

{  Our choice of the DM halo masses is based on a number of factors: (1) avoiding low-mass galaxies which cannot form sustainable galactic disks on one hand, and avoiding the high mass halos and galaxies which would be exceedingly rare at these redshifts, on the other hand; (2) Similar mass halos at different redshifts grow at different rates --- an effect which is expected to propagate down to galactic scales and which we plan to analyze; (3) Focusing on similar mass halos at different redshifts allows us to single out two processes, the effect of environment and that of the stellar feedback; (4) Moreover, such DM halos are expected to host galaxies with stellar masses of and in excess of $10^{10}\,{\rm M_\odot}$ being most hospitable to developing stellar bars as observed in the contemporary universe \citep[e.g.,][]{gad09}; (5) Most importantly, these halos growth is expected to contain $L*$ galaxies at $z=0$, so their comparison with the present day galaxies is the most optimal one. }

{  To some extent, we are still testing the simplest scenario of galactic structure formation when the disk component is produced by a diffuse accretion and the stellar spheroid origin results from the mostly dissipationless mergers in the hierarchical universe \citep[e.g.,][]{white78}.  Yet, some significant details of this scenario are still elusive, and we are unable to provide a quantitative explanation to the Hubble Fork diagram in the contemporary universe \citep[e.g.,][]{scanna09,roma09,crain15,schaye15,vogel18a,vogel18b,pille19}.}

{  Numerical simulations have been indispensable in understanding the formation of realistic galaxies and determining the emerging morphologies, using both large computational boxes to quantify statistical properties of samples of galaxies, and by means of the zoom-in simulations of individual objects. The large box simulations have allowed to focus on the evolution of scaling relations, but their mass and spatial resolution are often insufficient to follow the dynamics of the inner region of galaxies \citep[e.g.,][]{vogel14,schaye15,nel18,nel19}. On the other hand, zoom-in simulated galaxies and halos can deal only with a small number of objects due to the associated heavy computational load \citep[e.g.,][]{nava94,spri08,roma09,gue11,pallo17}. Modeling galaxy evolution at high redshifts is challenging, as it requires substantially higher mass and spatial resolutions. However, insufficient resolution at high redshifts can result in a skewed evolution of galaxies at low redshifts, including the effect of stellar bars on this evolution. }

Our suite of high-resolution zoom-in cosmological simulations explores different ways to characterise the morphologies, which for various reasons are not observed at low redshifts. This exploratory work attempts a number of options to quantify and compare the set of parameters which define the galactic  morphology at $\sim 1-3$\,Gyr after the Big Bang. We follow the assembly of DM halos, which are not the {\it most} massive at these redshifts. Halos with this mass range are smaller than the Milky Way halo only by a factor of 2.5 at $z=0$ \citep[e.g.,][]{posti19}. They are not common at redshifts 2, and become progressively rare towards $z_{\rm f}=6$.
 
In the local universe, observations have detected a trend towards an organized galactic rotation above the stellar masses of ${\rm log}\,M_*\sim 9 - 9.5$\,M$_\odot$ \citep[e.g.,][]{kass12,simo15,simo17,badry18,pille19}, thus reflecting transition to the Tully-Fisher (TF) relation \citep[e.g.,][]{tully77}. Lower mass galaxies show progressively increasing scatter in the TF diagram, a more diverse morphology and kinematics. This galaxy population exhibits irregular morphology and velocity fields, including those of the gas component \citep[e.g.,][]{walt08,ott12}. That the prevailing galactic morphology is changing with redshift has been detected clearly already at $z\sim 0.6$, when the fraction of peculiar galaxies has grown by a factor of $\sim 2-3$ on behalf of disk galaxies \citep[e.g.,][]{delga10}, and it is expected to grow even further with $z$ \citep[e.g.,][]{brinch00}.

While it seems logical that the mergers and flybys play an important role in shaping the galaxies, development of galactic morphologies is full of controversies. At the face value, frequent mergers could disrupt fragile galactic disks and one should expect the prevalence of spheroids at high redshifts. However, the cold accretion flows can compete successfully with mergers driving the galaxy growth \citep[e.g.,][]{keres05,dekel06}. Numerical simulations have also indicated that galaxies at $z\sim 9-10$ are preferentially disky, above galaxy masses of $\sim 10^9$\,M$_\odot$, in agreement with galaxies at $z=0$, which are sufficiently resolved \citep[e.g.,][]{roma11}, and grow preferentially by cold accretion flows in a broad range of masses above $z\sim 6$ at least \citep[e.g.,][]{roma14}. This has been extended to lower redshifts, although observationally only a circumstantial evidence exists for smooth accretion flows penetrating the DM halos \citep[e.g.,][]{sanci08,keres09,rau11,voort12,mart15, ver17}. For $z\ltorder 1$, the cold accretion flow should decrease sharply \citep[e.g.,][]{brooks09}.

{  Understanding galactic morphology presumes that one is capable of separating the stellar body of the galaxy into subcomponents based on the surface and/or brightness density, kinematics, metallicity, age, etc.  Attempts to decompose galaxies into various morphological features have a long history and aims at quantifying their relative importance during galaxy evolution \citep[e.g.,][]{abadi03a,abadi03b,gover07,dimatt08,croft09,mari14,genel15,jag21}. }

In this work, {  we focus on quantifying the galactic morphology by various methods, namely, applying the bulge-disk decomposition using the surface stellar density, the surface photometry, and the stellar kinematics. We compare the numerical images of our galaxies with their synthetic images using the James Webb Space Telecsope (JWST) and  Atacama Large Millimeter/submillimeter Array (ALMA). We also perform the bulge-disk decomposition for these photometric images for comparison with our numerical models.} 

{  The kinematic decomposition of a stellar system used in the literature is based on the method advanced by \citet{abadi03b} and typically applied only to $z=0$ objects. It is based on a single parameter of a normalized specific stellar angular momentum, and assumes that the spheroidal component has no angular momentum, with an equal amount of stars on the prograde and retrograde orbits, as well as the presence of thin and thick disk components (in some publications) which have different rotational support. This separation into cold and hot disk components is completely arbitrary. We apply a different decomposition which uses a single disk component but separates the bulge from the extended and hotter stellar spheroid, without restricting their angular momenta. Contrary to expectations of "cold" disk with the aspect ratios of 10:1 as mentioned in the literature for $z=0$ disks, our disks at $\gtorder 2$ are expected to appear hotter and with smaller aspect ratios.  We also do not expect that the spheroid is spherical.}

{  A modified method for kinematic decomposition disk -- spheroid based on two parameters has been applied by  \citet[][]{scanna09}. We have tested this method and an additional method based on two parameters, and discuss the results in section\,\ref{sec:kinematics}. }

While we do not anticipate that each method returns identical result, some correspondence is expected. In particular, we ask the following questions: how do global and local parameters describing the galaxy assembly correlate among similar mass parent DM halos at various $z_{\rm f}$. By global parameters we mean the stellar and baryonic masses as well as galaxy sizes, and overall rotational versus dispersion-supported motions in stellar and gaseous components in galaxies. By local parameters we mean the bulge-to-total ratio of stellar masses, B/T. Are bulges dynamically cold (disky) or hot (classical), i.e., are they rotationally or dispersion supported? We employ two types of stellar feedback --- galactic winds of a different strength, and anlyze how does the wind's strength affect the galaxy evolution at these redshifts? Furthermore, we follow the stellar mass buildup of galaxies and trace the stellar migrations from their birth place to that at $z_{\rm f}$. 

{  The galaxy growth rate in similar mass halos at different redshifts is expected to be influenced by a substantially different parent DM halo growth, even when immersed in the same overdensity environment. The stellar feedback, either in the form of SN or galactic winds, is expected to scale with the star formation rate, and do this in nonlinear fashion. We analyze the effect this growth rate has on the galactic morphology --- to the best of our knowledge this method was not used so far in the literature. } 

In short, we are interested to compare the structural and kinematic evolution of  high-$z$ galaxies that are hosted by similar mass halos subject to different stellar feedback at different redshifts. In a subsequent paper, we analyze the properties of galactic bars formed in these galaxies \citep[][hereafter paper\,II]{bi21}.

This paper is organized as follows. Section\,2 describes the numerical issues and methods used. Section\,3 presents our results, including photometric and kinematic analysis of bulges and galaxies, and population evolution and migration in these objects. This is followed by discussion section and summary.

%%%%%%%%%%%%%%%%%%%%%%%%%%%%%
\section{Numerical Modeling}
\label{sec:modeling}
%%%%%%%%%%%%%%%%%%%%%%%%%%%%%

%%%%%%%%%%%%%%%%%%%%%%%%%%%%%%%%%%%%%%%%%%%%
\subsection{Numerics and Initial Conditions}
\label{sec:numerics}
%%%%%%%%%%%%%%%%%%%%%%%%%%%%%%%%%%%%%%%%%%%%

\begin{table*}
%\caption{  Simulation suite with different galactic wind feedback, constant wind (CW) and variable wind %(VW).}.  
\vspace*{-.1cm}
%\title{  Preliminary Sample Estimates}
\centering
%\resizebox{\columnwidth}{!}{%
\begin{tabular}{ccccccccccc}
\hline
 $\mathrm{z_{\rm f}}$ & Model Name & ${\rm log}M_{\rm vir}\,$M$_\odot$ & ${\rm log}\,M_{\rm vir}\,$M$_\odot$ & $R_{\rm vir}\,{\rm kpc}$ & $R_{\rm vir}\,{\rm kpc}$ & $\uplambda_{\rm DM}$ & $\uplambda$ & $\updelta$ & Feedback \\
          & & DM only sims. & baryonic sims. & DM only sims. & baryonic sims. & DM only sims.& baryonic sims.    &     &       \\
\hline
\hline   
    6     & Z6HCW & 11.7  & 11.6 & 263 & 252 & 0.039& 0.024  & 3.04 & CW  \\
    6     & Z6HVW & 11.7  & 11.6 & 263 & 257 & 0.039& 0.024  & 3.04 & VW  \\
\hline   
    6     & Z6LCW & 11.7  & 11.6 & 263 & 251 & 0.028& 0.023  & 1.60 & CW  \\
    6     & Z6LVW & 11.7  & 11.6 & 263 & 249 & 0.028& 0.018  & 1.60 & VW  \\
\hline 
\hline   
    4     & Z4HCW & 11.7  & 11.6 & 263 & 271 & 0.035& 0.061  & 3.00 & CW  \\
    4     & Z4HVW & 11.7  & 11.6 & 263 & 254 & 0.035& 0.055  & 3.00 & VW  \\
\hline   
    4     & Z4LCW & 11.8  & 11.6 & 264 & 258 & 0.019& 0.022  & 1.33 & CW  \\
    4     & Z4LVW & 11.8  & 11.6 & 264 & 267 & 0.019& 0.024  & 1.33 & VW  \\
\hline
\hline   
    2     & Z2HCW & 11.8  & 11.7 & 295 & 293 & 0.019& 0.023  & 2.80 & CW  \\   
    2     & Z2HVW & 11.8  & 11.7 & 295 & 293 & 0.019& 0.054  & 2.80 & VW  \\ 
\hline      
    2     & Z2LCW & 11.8  & 11.7 & 279 & 283 & 0.020& 0.021  & 1.47 & CW  \\    
    2     & Z2LVW & 11.8  & 11.7 & 279 & 261 & 0.020& 0.013  & 1.47 & VW  \\
  
\hline   
\hline
\end{tabular}
%}
\caption{Simulation suite with DM only and with baryons. All values are given at the final redshifts $z_{\rm f} = 6$, 4, and 2. The columns correspond to (from left to right): the final redshift $z_{\rm f}$; the model number (see definition in section\,{sec:group}; the virial mass of DM halo $M_{\rm vir}$ at $z_{\rm f}$ in DM only and baryonic simulations; the halo virial radius (in comoving coordinates) $R_{\rm vir}$ in DM only and baryonic simulations; $\uplambda$ --- the DM halo spin in DM only and baryonic simulations; $\updelta$ --- the local overdensity; CW and VW --- the galactic wind feedback, Constant Wind (CW) and Variable Wind (VW).} 
\label{tab:DMsim}
\end{table*}

We invoke our version of the hybrid $N$-body/hydro code \textsc{gizmo} \citep{hopk17} with the Lagrangian meshless finite mass (MFM) hydro solver. \textsc{gizmo} has implemented an improved version of the TreePM gravity solver from \textsc{gadget} \citep{spri05}.  

We assume the Planck\,16 $\Lambda$CDM concordant cosmology with parameters $\Omega_{\rm m} = 0.308$, $\Omega_\Lambda = 0.692$, $\Omega_{\rm b} = 0.048$, $\sigma_8 = 0.82$, and $n_{\rm s} = 0.97$ \citep{planck16}. The Hubble constant is taken as $h = 0.678$ in units of $100\,{\rm km\,s^{-1}\,Mpc^{-1}}$. The  initial conditions have been imposed at the starting $z=99$. Both the parent box and the individual zoom-in simulations were created using the code \textsc{music} \citep{hahn11}. The simulations have been run from $z = 99$ to the target redshifts $z_{\rm f} = 6$, 4, and 2. Additional details of the models are given in Table\,\ref{tab:DMsim}. 

The meshless finite mass (MFM) hydrosolver and an adaptive gravitational softening for the gas has been implemented. The multiphase interstellar matter (ISM) algorithm has been used citep{spri03}, and the stellar feedback included the SN\,II and two models of galactic winds (see section\,\ref{sec:winds}). Feedback by SN\,Ia appears to be unimportant for $z\gtorder 2$ \citep[e.g.,][]{child14}, and therefore was not included. Simulations included the redshift-dependent cosmic UV background \citep{Faucher09}. 

{  We use the hybrid multiphase model for the ISM and star formation \citep{spri03}, where starforming particles contain the cold phase that forms stars and the hot phase that results from the SN heating. The former contributes to the star formation, while the latter contributes to the thermal gas pressure. This recipe includes the metal enrichment --- the metallicity increase in the starforming gas is related to the fraction of gas in the cold phase, the fraction of stars that turn into SN, and the metal yield per SN. We follow the gas and stellar metallicities, including H, He, C, N, O, Ne, Mg, Si, S, Ca, and Fe.    Metal diffusion is not implemented explicitly, but metals can be transported by winds (see section\,\ref{sec:winds}). The density threshold for star formation (SF) was set to $n^{\rm SF}_{\rm crit}=4\,{\rm cm^{-3}}$.}  

The gas cooling, including Compton, free-free, collisional, and metals, and heating processes are accounted for, including effects of ionization and recombination.   

{  For the zoom-in part, the mass per particle is $3.5\times 10^4\,{\rm M_\odot}$ (for gas and stars) and $2.3\times 10^5\,{\rm M_\odot}$ (for DM). In comoving coordinates, the minimal adaptive gravitational softening for the gas was 74\,pc, and the softening for stars and DM is 74\,pc and 118\,pc. This means, for example, that at the final redshifts, $z_{\rm f}=6$, 4, and 2, the softening for the stars in physical coordinates is 10.5\,pc, 14.7\,pc, and 24.6\,pc, respectively. The effective number of baryonic particles in our simulations is $2\times 4096^3$. 

The Illustris-1, IllustrisTNG100-1 and TNG50 simulations \citep{vogel14,nel18,nel19} have the DM mass per particle $6.3\times 10^6\,{\rm M_\odot}$, $7.5\times 10^6\,{\rm M_\odot}$, and $4.5\times 10^5\,{\rm M_\odot}$, respectively. The gas and stellar mass per particle is $1.6 \times 10^6\,{\rm M_\odot}$, $1.4\times 10^6\,{\rm M_\odot}$ and $8.5\times 10^4\,{\rm M_\odot}$. The EAGLE simulations \citep{schaye15} have the mass resolution of $9.7\times 10^6\,{\rm M_\odot}$ for DM and  $1.8\times 10^6\,{\rm M_\odot}$ for the gas. The gravitational softening in Illustris-1, TNG100-1 and TNG50 is 1.4\,kpc, 0.74\,kpc and 0.29\,kpc for DM,  and 0.74\,kpc, 0.19\,kpc and 0.074\,kpc for the gas, respectively. In EAGLE simulations the resolution is 0.7\,kpc at $z=2.8$. This softening is depending on redshift for $z > 1$. Hence, with the exception of the TNG50 simulations, both the mass and spatial resolutions are substantially reduced in comparison to those in the present work, up to two orders of magnitude. The TNG50 is compatible to our simulations within a factor of 2--3.}

%%%%%%%%%%%%%%%%%%%%%%%%%%%%%%%%%%%%%%%%%%%%%
\subsection{Defining DM halos and galaxies}
\label{sec:group}
%%%%%%%%%%%%%%%%%%%%%%%%%%%%%%%%%%%%%%%%%%%%%

The algorithm \textsc{music} \citep{hahn11} has been used to initial conditions for the parent box and the individual zoom-in simulations. \textsc{music} uses a real-space convolution approach in conjunction with an adaptive multi-grid Poisson solver to form highly accurate nested density, particle displacement, and velocity fields suitable for multi-scale zoom-in simulations of structure formation in the universe. Firstly, we have generated a comoving box of 74\,Mpc with $1024^3$ DM-only particles, and run it until redshift $z=2$. The DM halos have been selected applying the group finder (see below) at the target redshifts of $z_{\rm f} = $ 6, 4, and 2, to have the prescribed DM masses of log\,($M_{\rm vir}/{\rm M_\odot})\sim 11.75\pm 0.05$, dimensionless halo spin $\uplambda$, and the local overdensity $\updelta$ (see below for the definitions of $M_{\rm vir}$ and $\updelta$). 

The halos have been calculated by using the \textsc{rockstar} group finder \citep{behr13}, with Friends-of-Friends (\textsc{FoF}) linking length of $b=0.28$, and accounting only for the bound DM particles (Table\,\ref{tab:DMsim}). The merger trees have been constructed using the \textsc{rockstar} halo catalogs and the \textsc{consistent-trees} algorithm \citep{behr12}. The halo virial radius  and the virial mass, $R_{\rm vir}$ and $M_{\rm vir}$, have been defined by $R_{200}$ and $M_{200}$ \citep[e.g.,][]{nfw96}. $R_{200}$ is the radius inside which the mean interior density is 200 times the critical density of the universe at that time. 

In the next step, we have carved out a sphere encompassing all particles within the volume of $\sim 3R_{\rm vir}$ to avoid contamination of the high resolution region by the massive particles. These particles have been traced to their initial conditions, in order to create a mask for \textsc{music}, where the resolution has been increased by 5 levels, i.e., from $2^7$ to $2^{12}$. Overall, we have 5 levels of refinement, analyzing only the highest level --- our mass and spatial resolution given below refer only to this level.

The full box run has defined the overdensity environment $\updelta$ by creating 1.5\,Mpc grids centered on DM halos and calculating the average DM densities inside these grids. The ratios between these densities and the average density of the universe provide $\updelta$. The DM halos have been chosen for low and high overdensities, $\updelta\sim 1$ and $\updelta\sim 3$, respectively. Table\,\ref{tab:DMsim} lists the DM halo models in our simulations and their main characteristics.
 
We used the group-finding algorithm \textsc{hop} \citep{eise98} to identify galaxies using the outer boundary threshold of baryonic density $10^{-2}\,n^{\rm SF}_{\rm crit}$, which ensures that both the host star-forming gas and the lower density non-star-forming gas are roughly bound to the galaxy \citep{roma14}. We use \textsc{hop} for this purpose in order not to impose a particular geometry on the galaxy. Note that this definition of galaxy differs from many used in the literature, which base the galaxy size on $0.1R_{\rm vir}$ \citep[e.g.,][]{mari14}.

%%%%%%%%%%%%%%%%%%%%%%%%%%%%%%%%%%
\subsection{Galactic Wind Models}
\label{sec:winds}
%%%%%%%%%%%%%%%%%%%%%%%%%%%%%%%%%%

There is a consensus that stellar feedback plays a major role in galaxy evolution. Absent or weak feedback results in a very efficient star formation, leading to compact galaxies populated with old stars, in contradiction to observations \citep[e.g.,][]{white78,gover07}. We have used two types of galactic feedback models in addition to the stellar SN Type\,II feedback: the Constant Wind \citep{spri03} and the Variable Wind \citep{oppi06}, CW and VW, respectively. Both wind models have been implemented by "decoupling" the wind particles from the gas particles. In other words, the wind particles do not interact hydrodynamically and move ballistically. The time period of decoupling depends on the shortest time between $10^6$\,yrs and decrease in the background gas density by a factor of 10.

In the CW models, the wind velocity has been taken constant, $v_{\rm w} = 484\,{\rm km\,s^{-1}}$ and the wind mass-loading factor is $\beta_{\rm w}\equiv \dot M_{\rm w}/\dot M_{\rm SF} = 2$ \citep{roma14}. Here, $\dot M_{\rm w}$ represents the mass loss in the wind, and $\dot M_{\rm SF}$ is the (mass) star formation rate. The wind is assumed to be isotropic. In the VW models, the wind velocity scales with the parent DM halo escape speed. The wind mass-loading factor is calculated based on the wind energy and its momentum. The wind orientation is isotropic as well.

These winds allow us to vary the feedback and determine the effect of this feedback on galaxy evolution. The accepted parameters of the winds prescribe that the VW will have a stronger effect on the gas dynamics by depositing larger kinetic energy there. Indeed, we have measured the ratio of kinetic energies of $E_{\rm CW}/E_{\rm VW}$, and find that this ratio is always less then unity, and typically ranges between 0.1 and 0.01, sometimes diving below this values. It varies by a factor up to $10^3$. The expected result is that the star formation rate will be smaller in the VW models, which indeed is the case (see Section\,\ref{sec:general}). Additional corollary of this evolution is that the gas fraction in VW models is larger than in CW ones. 
 
{  We have avoided using the top-of-the line algorithms for stellar feedback, e.g., FIRE-2 \citep{hopk18}, because of two reasons. Namely, (1) the FIRE method becomes only realistic when specific spatial scales of energy and momentum deposition are resolved, i.e., few to few$\times 10$ pc \citep{mura15}, which is still slightly below our resolution limit. (2) Our goal is only to compare the effect of the weak and strong feedback on the basic galaxy parameters, including their morphology.}  
 
Tables\,\ref{tab:DMsim}, \ref{tab:deco_dens}, and \ref{tab:deco_photo} provide parameters of modeled galaxies at $z_{\rm f}$. The naming convention for the models is as follows: the galaxy model abbreviated as Z4HCW corresponds to the galaxy with the final redshift $z_{\rm f}=4$, high overdensity, and CW feedback. Galaxy model abbreviated with Z4LVW corresponds to the galaxy with the final redshift $z_{\rm f}=4$, low overdensity, and VW feedback.  
 
%%%%%%%%%%%%%%%%%%%%%%%%%%%%%%%%%%%%%%%%%%%%%%%%%%%%%%%%%%%%%%%%%%%%%%%%%%
\subsection{From stellar particles to galaxy light: the JWST and ALMA imaging}
\label{sec:light}
%%%%%%%%%%%%%%%%%%%%%%%%%%%%%%%%%%%%%%%%%%%%%%%%%%%%%%%%%%%%%%%%%%%%%%%%%%

In order to determine the observational properties of our galaxies, we have post-processed these galaxies with the 3-D radiation transfer code \textsc{SKIRT} \citep{baes03}, then redshifting them but not pixelizing, thus producing synthetic images of these galaxies face-on and edge-on. Each stellar particle was treated as a coeval single stellar population that has the \citep{chab03} initial mass function (IMF) and an assigned Bruzual-Charlot \citep{bruz93} spectral energy distribution (SED) family. The dust density has been assumed to be proportional to the gas metallicity with the coefficient of proportionality 0.3 \citep[e.g.,][]{camps16}. The total emission has included the secondary emission from the dust and iterations for dust self-absorption. With these SEDs, we decided which of the James Webb Space Telescope (JWST) filters should be used at different redshifts, in order to obtain the brightest images in observational frames. Finally, we obtained the observational images with the NIRCam at $z_{\rm f}=6$ with the three-color mosaic of F070W, F115W and F365W filters, and the images at $z_{\rm f}=4$ \& 2 with the three-color mosaic of F070W, F115W and F200W filters (see Fig.\,\ref{fig:photo_images}). 

Similarly, galaxies with $z_{\rm f}$ have been also processed for the Atacama Large Millimeter/submillimeter Array (ALMA) imaging, by using ALMA Observational Support Tool (OST). Again, we used \textsc{SKIRT} to generate the source flux profile at a central frequency 93.7\,GHz and a bandwidth of 7.5\,GHz. The Cycle\,7 array configurations with the maximal baseline of 16,197m and the lowest level (1st octile) of precipitable water vapour (PWV = 0.472\,mm) have been applied to simulate the observational ALMA images (see Fig.\,\ref{fig:ALMA}).  

%%%%%%%%%%%%%%%%%%%
\section{Results}
\label{sec:results}
%%%%%%%%%%%%%%%%%%%

\begin{figure}
%\center 
 \includegraphics[width=0.48\textwidth]{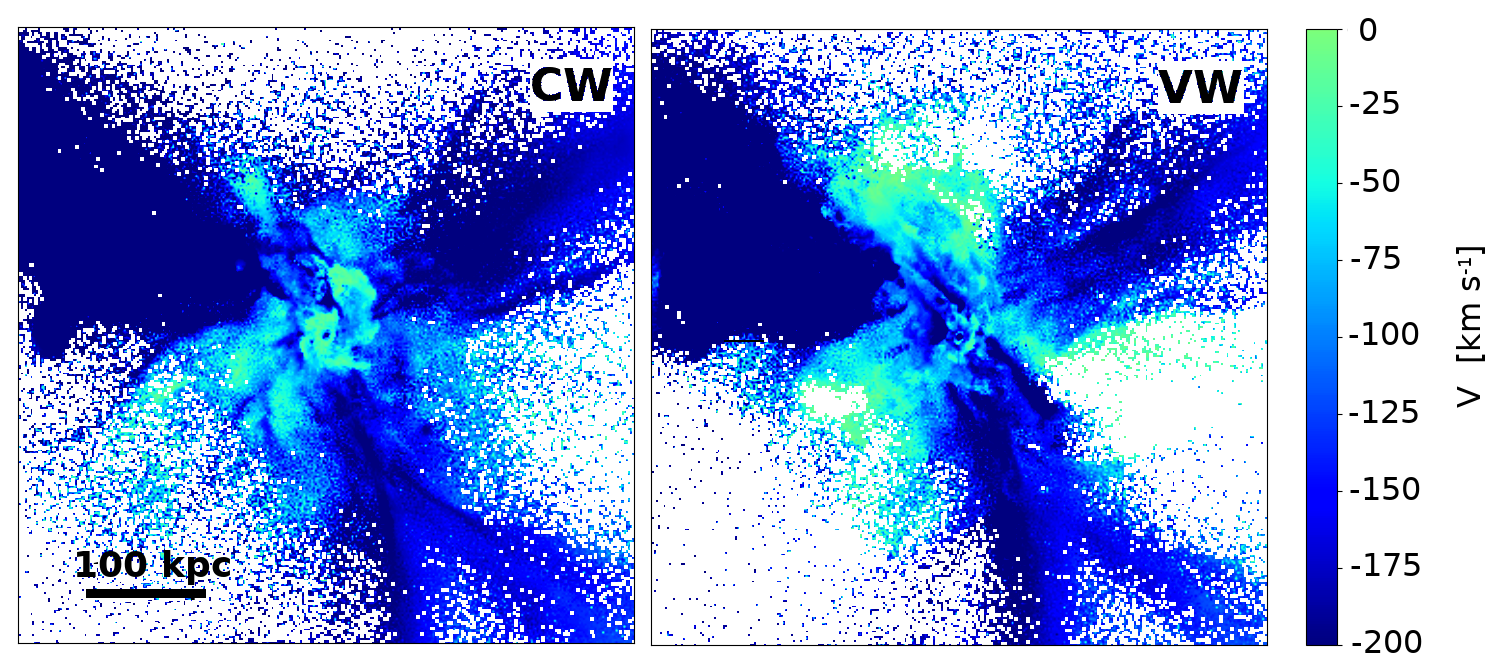}
    \caption{Examples of cold gas accretion flows at $z = 2.28$ in CW (left) and VW (right) for the DM halo Z2L with $z_{\rm f} = 2$. The color palette reflects only infalling radial velocity of the gas. The frame size is $500\,{\rm kpc}$ in comoving coordinates. The difference in the central region is due to the stellar feedback which is stronger in the VW models.}
    \label{fig:inflow}
    \end{figure} 

\begin{figure}
%\center 
\includegraphics[width=0.49\textwidth]{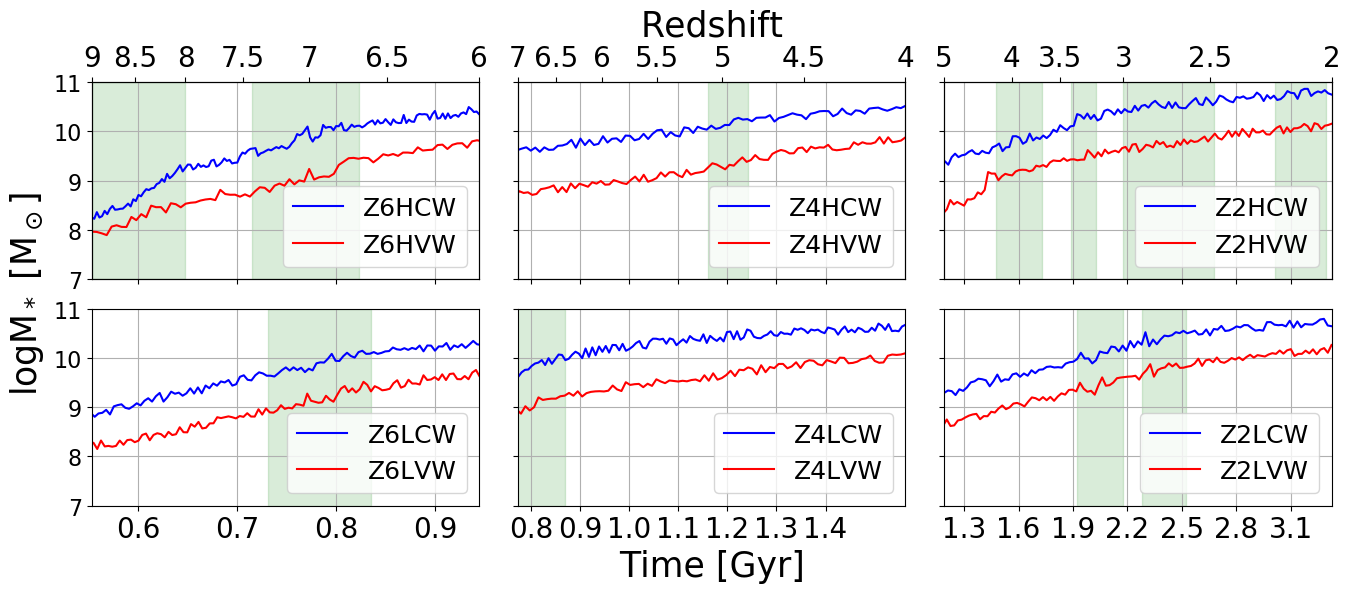}
    \caption{Evolution of stellar masses in galaxies, in physical coordinates. The VW models are given by the red lines, and CW models by the blue lines. The green colored background refers to major merger events. Note that the minor and intermediate mergers, and close flybys, as well as periods of massive cold accretion inflows can have equally strong effect on the galaxy properties, but are not marked for clarity. The top row here and elsewhere in the subsequent figures represents galaxies in the high overdensity regions, while the bottom row displays galaxies in the low overdensity regions. }
    \label{fig:Smass}
    \end{figure} 
    
\begin{figure}
\center 
	\includegraphics[width=0.49\textwidth]{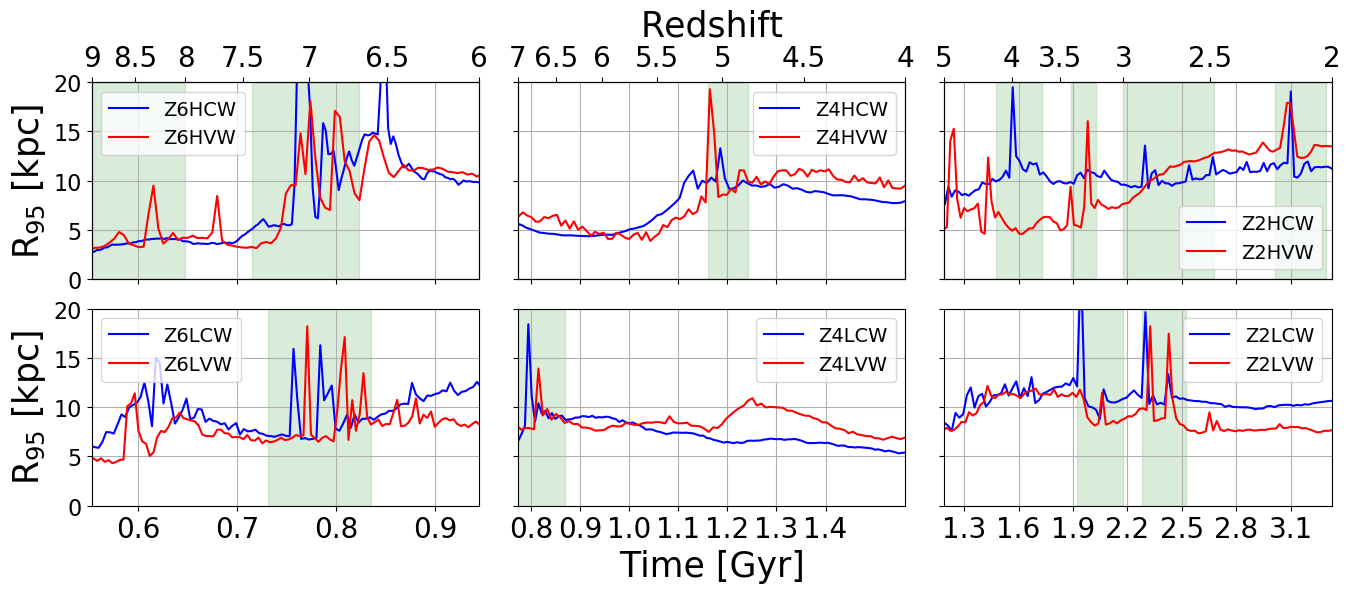}
    \caption{  Evolution of galaxy sizes containing 95\% of the stellar mass defined by the \textsc{hop} in comoving coordinates.  The VW models are given by the red lines, and CW models by the blue lines. The green colored background refers to major merger events. }
    \label{fig:radii}
    \end{figure}      
    
\begin{figure}
%\center 
\includegraphics[width=0.49\textwidth]{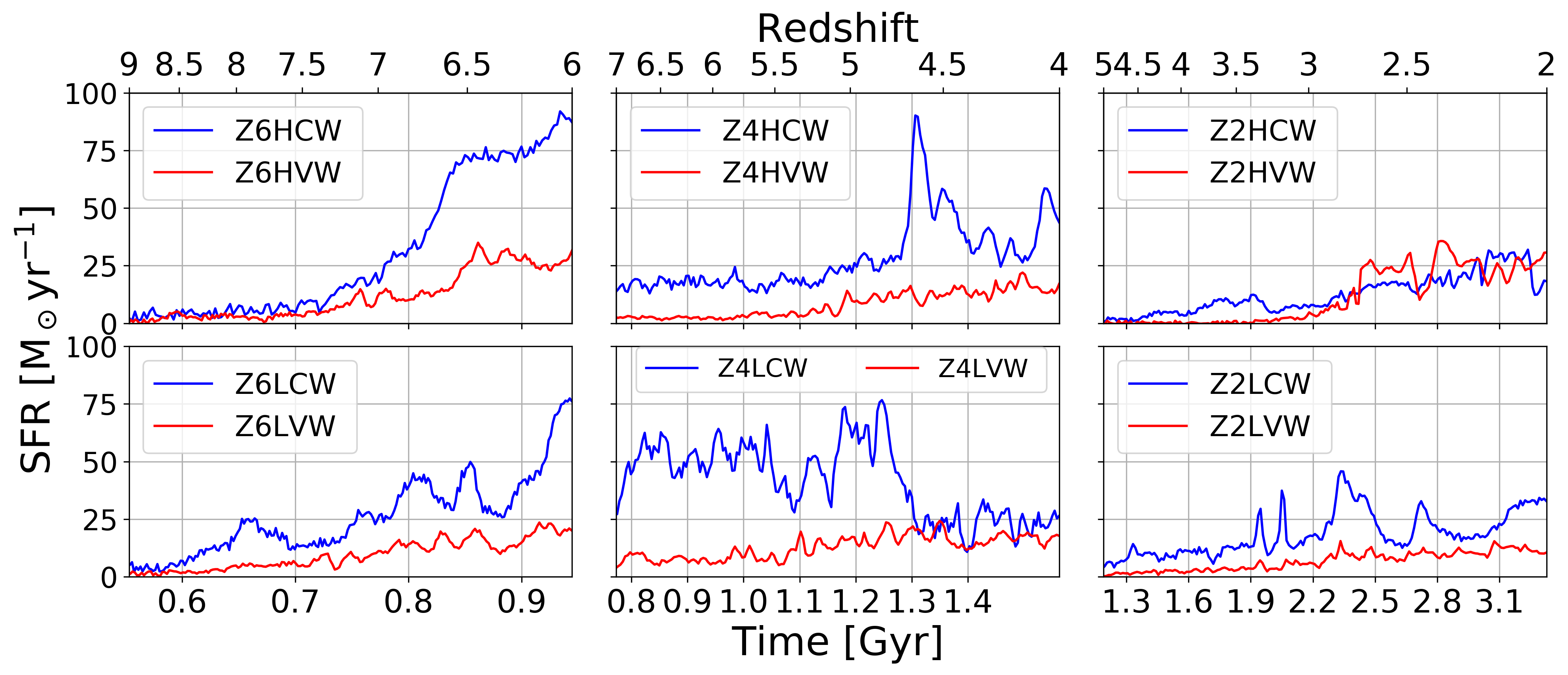}
    \caption{Evolution of the SFR in galaxies. The VW models are given by the red lines, and CW models by the blue lines.}
    \label{fig:galSFR}
    \end{figure}     

 \begin{figure}
\center 
	\includegraphics[width=0.49\textwidth]{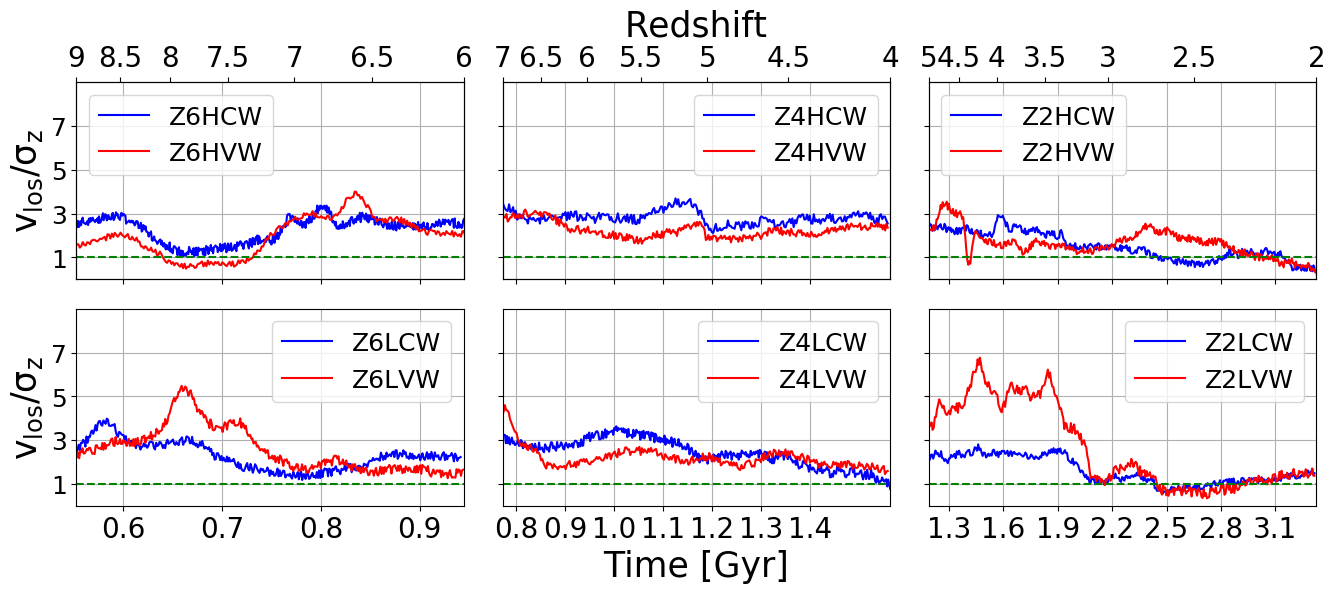}
    \caption{The ratio $v_{\rm los}/\sigma_{\rm z}$, line-of-sight velocity-to-vertical dispersion velocity of stars, as a function of time, for all galaxy models  up to their respective final redshifts, $z_{\rm f}=6, 4,$ and 2, for CW (blue lines) and VW (red lines). The dashed line corresponds to $v_{\rm los}/\sigma_{\rm z}=1$. For $\sigma_{\rm z}$, the average velocity has been used, while for $v_{\rm los}$ we use the maximal values.}
    \label{fig:vsigma_stars}
    \end{figure} 

 \begin{figure}
\center 
	\includegraphics[width=0.49\textwidth]{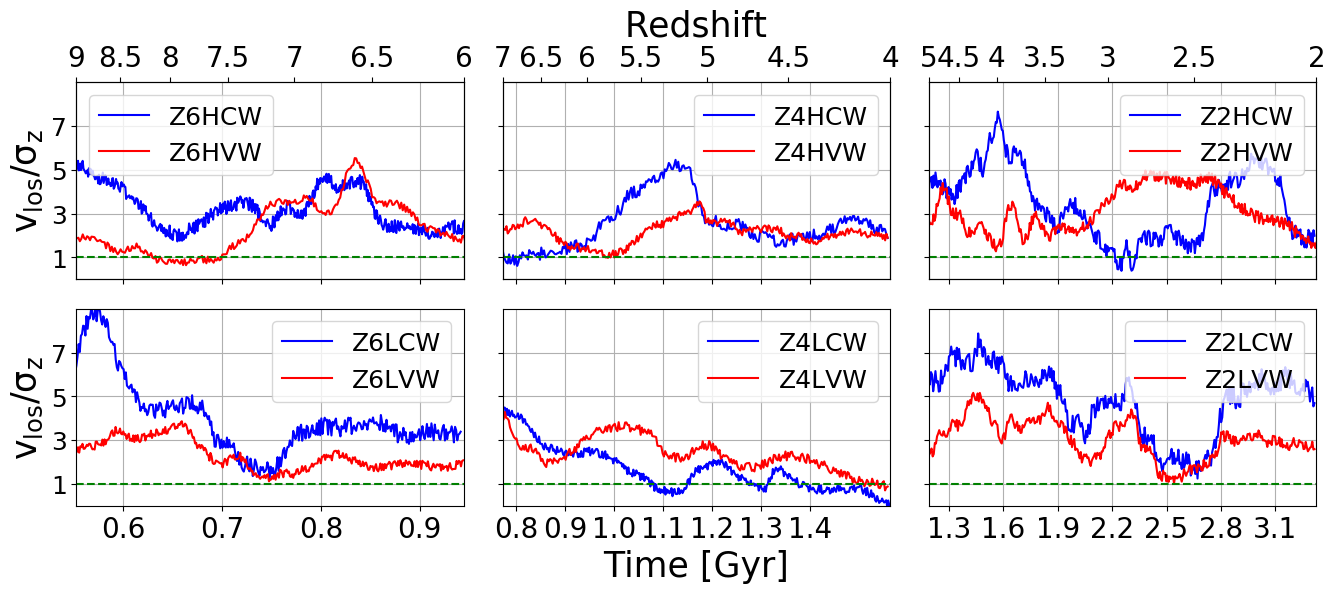}
    \caption{The ratio $v_{\rm los}/\sigma_{\rm z}$, line-of-sight velocity-to-vertical dispersion velocity of the gas, as a function of time, for all galaxy models up to their  final redshifts, $z_{\rm f}=6, 4,$ and 2, for CW (blue lines) and VW (red lines).  The dashed line corresponds to $v_{\rm los}/\sigma_{\rm z}=1$. For $\sigma_{\rm z}$, the average velocity has been used, while for $v_{\rm los}$ we use the maximal values.}
    \label{fig:vsigma_gas}
    \end{figure} 
  
 To understand the evolution of the central galaxies in our models, one should remember that the parent DM halos have been chosen to have similar masses at specific redshifts, i.e, $z_{\rm f}=6$, 4 and 2. We limit the redshifts shown for galaxy evolution to $z=9-6$, $7-4$ and $5-2$, to roughly include the time periods when the galaxies are well resolved numerically. The DM halos final masses in baryonic simulations lie within the virial masses of log\,$M_{\rm vir}\sim 11.65\pm 0.05\,{\rm M_\odot}$. The corresponding halo sizes fall within $R_{\rm vir}\sim 250-290$\,kpc in comoving coordinates. 
    
 \begin{figure}
\center 
	\includegraphics[width=0.49\textwidth]{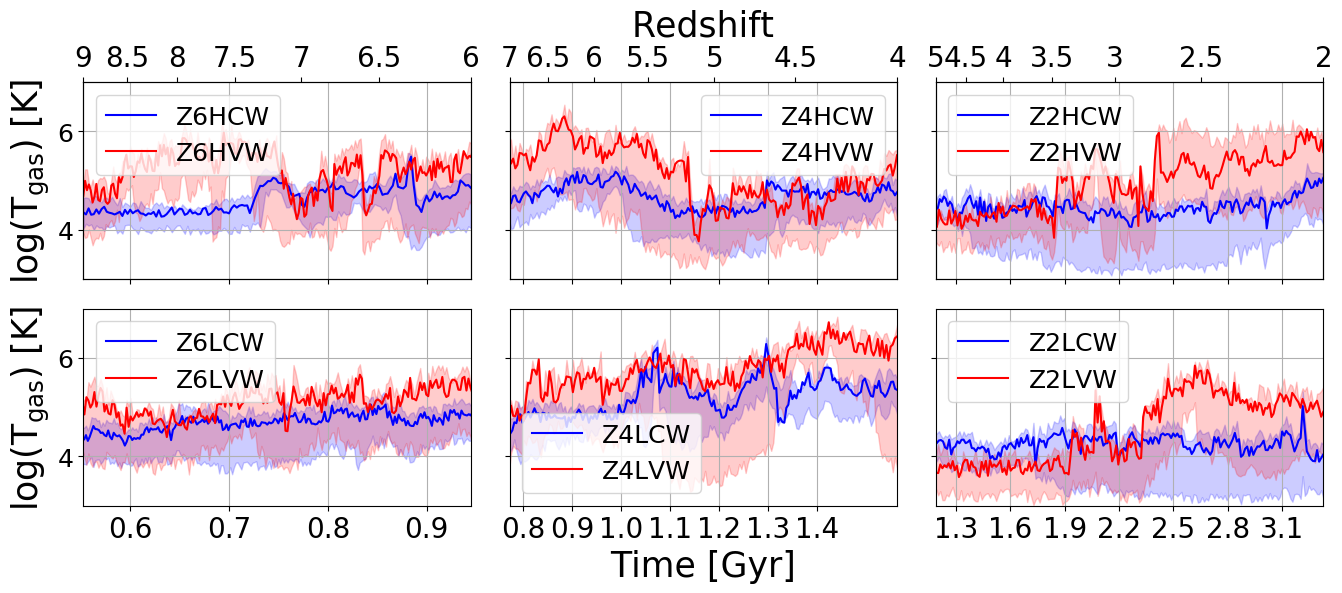}
    \caption{Evolution of the gas temperature inside galaxies.  The VW models are given by the red lines, and CW models by the blue lines. The color shadows show the 20-80 percentile range.}
    \label{fig:galgasT}
    \end{figure}    

Due to the selection process of the DM halos, our galaxies form and evolve in very different epochs, yet their global parameters, e.g., masses and sizes, are found within a relatively narrow range at the chosen $z_{\rm f}$. The total baryonic galaxy masses lie within the range of $3-7\times 10^{10}\,{\rm M_\odot}$ (Fig.\,\ref{fig:Smass}), and their sizes in the range of $8-12$\,kpc in comoving coordinates (Fig.\,\ref{fig:radii}). As we show below, this similarity does not propagate down to detailed internal evolution of galaxies, whose morphology, kinematics and stellar populations evolve in a very diverse way, and so are the final products of this evolution, as shown in Table\,\ref{tab:deco_dens}.

The simplest approach to understand the galactic morphology is to separate it into spheroidal and disky components. We proceed along this line and perform decomposition based on the stellar surface density first, followed by decomposition based on the photometric images of the same galaxies, redshifted and pixelized using the JWST specifications. Moreover, we use a bulge-disk and a spheroid-disk decompositions. As a next step, we analyze morphology based on the stellar kinematics. We provide the realistic images of our galaxies with the JWST and with ALMA. Lastly, we pay attention to evolution of stellar populations and the history of their migration in the host galaxies and their parent halos. 

Following the evolution of high-$z$ galaxies, one expects a substantial contribution of a number of factors, namely, of cold accretion flows, mergers. flybys, and interactions with the DM component. As an example, we show the associated accretion for Z2LCW and Z2LVW halos at $z = 2.28$ (Fig\,\ref{fig:inflow}). The halo lies at the intersection of three filaments and at a relatively low overdensity of $\updelta\sim 1.47$, confirming that at this redshift the cold accretion is prominent. Away from the central region, the filaments are prominent and nearly identical between the two models. Closer to the center the feedback plays a role, and differences are clear --- the radial velocity of accreting matter is smaller in the VW model, especially along the rays which do not intersect the filaments. 

Sites of star formation in galaxies delineate the underlying morphology. These sites are heavily biased towards the central regions. In addition, the disk growth amplified by mergers and cold accretion is associated with spiral arms which dominate the outer disks, both in gas and in stars. These spiral arms are gas rich and fragment forming stars, become clumpy and form stellar clumps. In our galaxies, this period corresponds to redshifts when the disk experiences non-axisymmetric perturbations, internal and external. During follow-up evolution, these clumps form stars and their gas component is gradually ablated. Many of these stellar clumps spiral-in with time and are disrupted in the central regions by $z_{\rm f}$. 

A major internal factor affecting morphology of disk galaxies is the presence of stellar bars. We do find stellar bars in our galaxies, but discuss them only briefly in this paper --- Paper\,II is dedicated to stellar bars. 

%%%%%%%%%%%%%%%%%%%%%%%%%%%%%%%%%%%%%%%%%
\subsection{Simulated galaxies: general}
\label{sec:general}
%%%%%%%%%%%%%%%%%%%%%%%%%%%%%%%%%%%%%%%%%

Figure\,\ref{fig:Smass} shows that all the CW galaxies have exceeded the stellar mass of $10^{10}\,{\rm M_\odot}$ by $z_{\rm f}$. The stellar components in VW galaxies are less massive by a factor of a few, because of a larger gas fraction and lie slightly below $10^{10}\,{\rm M_\odot}$ at $z_{\rm f}$. Both show no dependency on $\delta$. The stellar mass growth appears to be monotonic, except during the major mergers and other catastrophic events, as we show later\footnote{We have marked the periods of major merger events on this figure. However, minor and intermediate mergers, close flybys, and sudden influxes of cold accretion can be as important for galaxy evolution. For brevity, we did not mark them, but discuss them in the text.}. As galaxies reach approximately the same baryonic masses at $z_{\rm f}$, the gas mass fraction, $f_{\rm gas}$, separates the CW from the VW sequence. For all the galaxies, we observe that the growth exhibits some degree of saturation towards respective $z_{\rm f}$, most prominent for $z_{\rm f} = 2$ galaxies.

The similarities and differences between the SFR evolution at different redshifts can be observed in Figure\,\ref{fig:galSFR}. First, the SFRs in CW galaxies are substantially higher than in the VW galaxies. It displays a strong variability in the SFRs but with steady increase with redshift. To correlate the variability of SF with (for example) periods of major mergers, one can refer to Figure\,\ref{fig:Smass} where these periods have been marked. The galaxies Z6HCW and Z6HVW, which have mostly a similar merger history, reveal that a sharp increase in the SFR happens immediately with the end of the major merger at $z\sim 6.7$, in both objects which differ only by stellar feedback. Of course, additional higher frequency variability can be noticed and traced by us to less catastrophic events. In some cases, it is difficult to disentangle the cause from the consequence in this evolution. But it generally helps to separate the internal causes from externals ones, such as effects of stellar bars from mergers, which is especially important in Paper\,II. 

Galaxy Z4LCW serves as a nice example of the contribution of other effects, besides mergers, play an important role. At z$\sim 4.8$, it has experienced a sharp decline in the SFR, while no major merger event is associated with this redshift. A more careful check reveals a number of details: the stellar spheroid-to-total stellar mass, S/T, experienced increase at the time, the pattern speed of the stellar bar has sharply increased, while the SFR in the bar has has decreased, the fraction of gas in the bar has decreased, and finally the gas temperature has increased  (these details can be observed in figures of this paper and in Paper\,II). The verified conclusion of this decline in the SFR is that the galaxy experienced a close prograde flyby which led to misalignment between the gaseous and stellar disks, a speed-up of the bar and a heating of the gas --- the overall result is a sharp decrease in the SFR. 

Next, we turn to stellar and gas kinematics, and estimate the rotational support in galaxies versus vertical dispersion velocities (Figs.\,\ref{fig:vsigma_stars} and \ref{fig:vsigma_gas}). First, we note that most of the time both stars and gas are rotationally supported, i.e., $v_{\rm los}/\sigma_{\rm z} > 1$, where $v_{\rm los}$ and $\sigma_{\rm z}$ being the line-of-sight rotational velocity (in edge-on galaxies) and vertical dispersion velocities, respectively. Yet occasionally the $v_{\rm los}/\sigma_{\rm z}$ curve dives to unity. The stellar and gas components display this ratio being larger for CW models compared to VW models, but sometimes this trend is inverted, e.g., for the stellar component in the Z2L halo. In this case, the sharp decrease in $v_{\rm los}/\sigma_{\rm z}$ at around $z\sim 3.5$ (Fig.\,\ref{fig:vsigma_stars}) corresponds to a sequence of major mergers (Fig.\,\ref{fig:Smass}) which heats up the stellar disk substantially and causes the gaseous disk to become inclined to the stellar disk (Fig.\,\ref{fig:theta}). This event can be seen also in the gas temperature (Fig.\,\ref{fig:galgasT}).

The gas component shows larger amplitude variability in $v_{\rm los}/\sigma_{\rm z}$ and larger rotational support than the stellar component. Also, the CW gas shows larger rotational support than the VW gas, which is related to the stellar feedback strength and the galaxy interaction history. We do not observe a similar trend for the stellar components, which display a roughly similar rotational support, with the difference being insignificant. 

For $z_{\rm f}=6$ galaxies, the CW gas displays substantially larger rotational support than in the VW galaxies. After $z\sim 8$, this CW rotational support declines from $v_{\rm los}/\sigma_{\rm z}\sim 5-7$, and levels off at $v_{\rm los}/\sigma_{\rm z}\sim 3$. In the VW models, this support either raises to $v_{\rm los}/\sigma_{\rm z}\sim 3$ or stays flat at $v_{\rm los}/\sigma_{\rm z}\sim 2$. The stellar $v_{\rm los}/\sigma_{\rm z}$ ratio behaves in a similar fashion, but in the range of $v_{\rm los}/\sigma_{\rm z}\sim 2-3$. 

The $z_{\rm f}=4$ stellar models either level off at $v_{\rm los}/\sigma_{\rm z}\sim 3$ or dive to $v_{\rm los}/\sigma_{\rm z}\sim 1$. Hence in the latter case, the stellar disks lose their rotational support. The gas components follow the stellar components in this case, i.e., either oscillate around $v_{\rm los}/\sigma_{\rm z}\sim 3$ (galaxies Z4HCW and Z4HVW), or decline gradually to $v_{\rm los}/\sigma_{\rm z}\sim 1$ (Z4LCW and Z4LVW).   

Stellar models with $z_{\rm f}=2$  show weaker oscillations and a steady decline to $v_{\rm los}/\sigma_{\rm z}\sim 1$ (e.g., Z2LCW and Z2LVW), and even somewhat below it (e.g., Z2HCW and Z2HVW). Upon inspection, this trend originates from the appearance of a visible stellar spheroid component that envelops the stellar disk. So, $z_{\rm f}=2$ galaxies have in total much less rotational support in stars, both in CW and VW models. Note, this can be misleading as we include the stellar spheroid in this estimate, unlike in section\,\ref{sec:kinematics}. On the other hand, the gas component displays large amplitude oscillations in $v_{\rm los}/\sigma_{\rm z}$, related to mergers and interactions, yet end up with a substantial rotational support in all models.

In summary, by $z_{\rm f}$, all galaxies are massive for their respective redshifts and host stellar and gaseous disks in agreement with \citet{roma11}. The stellar and gaseous disks in our simulations exist most of the time shown in Figures\,\ref{fig:vsigma_stars} and \ref{fig:vsigma_gas}, i.e., starting with $M_*\gtorder 10^8\,{\rm M_\odot}$. In the $z_{\rm f}= 2$ galaxies, these disks coexist with growing stellar spheroids. We also see no evidence that gas becomes more rotationally supported with increasing baryonic mass, and find no importance for a characteristic mass of $10^{10}\,{\rm M_\odot}$, as observed in simulations with the \textsc{FIRE2} \citep[e.g.,][]{badry18}. We do observe that some massive stellar disks have gaseous component that have large dispersion velocities, but overall are still supported rotationally. These large dispersion velocities in the gas can be traced to a substantial warping in the disk plane, as we always define the galactic plane based on the stellar component. 

Comparing this evolution to low redshift galaxies, we do not detect an increased rotational support with increasing stellar or baryonic galaxy mass. On the contrary, galaxies with $z_{\rm f}=4$ and 2 show a somewhat declining rotational support with redshift. This trend can be traced to a number of factors, e.g., mergers, warped stellar disks, monotonically increasing SFR associated with $z\gtorder 2$ universe, and the anticipated increase in the stellar feedback acting on the gas. The stellar rotational support also reflects the substantial turbulent velocities in the star forming gas. The opposite trend which is observed in low redshift galaxies probably originates only at $z < 2$.

%%%%%%%%%%%%%%%%%%%%%%%%%%%%%%%%%%%%%%%%%%%%%%%%%%%%%%%
\subsection{Disk-bulge decomposition: surface stellar density}
\label{sec:density}
%%%%%%%%%%%%%%%%%%%%%%%%%%%%%%%%%%%%%%%%%%%%%%%%%%%%%%%

\begin{table*}
%\caption{  Simulation suite with different galactic wind feedback, constant wind (CW) and variable wind %(VW).}.  
\vspace*{-.1cm}
%\title{  Preliminary Sample Estimates}
\centering
%\resizebox{\columnwidth}{!}{%
\begin{tabular}{cccccccccc}
\hline
 $\mathrm{z_{\rm f}}$ & Model Name & $n_{\rm bulge}$ & $M_*{\rm (bulge)}\,{\rm M_\odot}$ & $R_{\rm e}{\rm (bulge)}$\,kpc & $M_*{\rm (disk)}\,{\rm M_\odot}$ & $R_{\rm e}{\rm (disk)}$\,kpc & $f_{\rm gas}$ & Feedback \\
\hline
\hline
    6 & Z6HCW & 2.27 & $1.32\times 10^{10}$ & 0.22 & $1.22\times 10^{10}$ & 2.48 & 0.50 & CW \\
    6 & Z6HVW & 0.54 & $0.31\times 10^{10}$ & 0.35 & $0.28\times 10^{10}$ & 3.20 & 0.87 & VW \\
\hline    
    6 & Z6LCW & 1.22 & $0.68\times 10^{10}$ & 0.21 & $1.29\times 10^{10}$ & 3.30 & 0.52 & CW \\
    6 & Z6LVW & 0.96 & $0.23\times 10^{10}$ & 0.31 & $0.25\times 10^{10}$ & 2.35 & 0.87 & VW \\
\hline
\hline
    4 & Z4HCW & 1.19 & $0.93\times 10^{10}$ & 0.27 & $2.28\times 10^{10}$ & 2.18 & 0.25 & CW \\
    4 & Z4HVW & 0.47 & $0.36\times 10^{10}$ & 0.35 & $0.36\times 10^{10}$ & 3.48 & 0.85 & VW \\
\hline
    4 & Z4LCW & 0.92 & $1.95\times 10^{10}$ & 0.31 & $1.35\times 10^{10}$ & 2.27 & 0.39 & CW \\
    4 & Z4LVW & 0.57 & $0.78\times 10^{10}$ & 0.35 & $0.27\times 10^{10}$ & 1.93 & 0.86 & VW \\

\hline
\hline
    2 & Z2HCW & 1.07 & $3.24\times 10^{10}$ & 0.50 & $3.37\times 10^{10}$ & 2.65 & 0.28 & CW \\   
    2 & Z2HVW & 0.50 & $2.57\times 10^{10}$ & 0.44 & $0.63\times 10^{10}$ & 5.46 & 0.77 & VW \\   
\hline
    2 & Z2LCW & 1.50 & $2.02\times 10^{10}$ & 0.47 & $3.31\times 10^{10}$ & 2.86 & 0.27 & CW \\
    2 & Z2LVW & 0.66 & $1.03\times 10^{10}$ & 0.37 & $0.49\times 10^{10}$ & 2.54 & 0.76 & VW \\

\hline   
\hline
\end{tabular}
%}
\caption{Stellar disk-bulge decomposition of modeled galaxies based on the surface stellar density at $z_{\rm f}$ in comoving coordinates. All models have been fitted with a bulge and an exponential disk components. From left to right: the final redshift $z_{\rm f}$; model name; the Sersic index of the bulge component $n_{\rm bulge}$; $M_{\rm bulge}$ and $R_{\rm e}{\rm (bulge)}$ --- the bulge stellar mass and the its half-stellar mass radius; $M_{\rm disk}$ and $R_{\rm e}{\rm (disk)}$ --- the disk stellar mass and its  half-stellar mass radius; $f_{\rm gas}$ --- the gas fraction in galaxies; B/T --- the bulge-to-total stellar mass ratio; CW and VW --- the type of the galactic wind feedback.} 
\label{tab:deco_dens}
\end{table*}

Results of stellar surface density decomposition into a bulge and an exponential disk are given in Table\,\ref{tab:deco_dens}, and in the Appendix, in Figures\,\ref{fig:BD_CWdens} and \ref{fig:BD_VWdens}. We note the following trends: the Sersic index for the bulge, $n_{\rm b}$, for all the CW models is larger than for the VW models, with a very small overlap. Specifically, $n_{\rm b} > 0.92$ for all the CW models, while $n_{\rm b} < 1$ for all the VW models. No dependence on $z_{\rm f}$ was found for the bulge index for the stellar surface density decomposition.

Next, the stellar bulge-to-total mass ratio, B/T, is larger for the VW models compared to the CW models. So, a VW bulge in comparison with the `twin' CW bulge is always more massive, despite that the parent galaxy has a larger gas fraction than the CW model, and clearly because of this. As all of our galaxies are barred most of the time, increased gas fraction leads to a larger gas accumulation and the stellar bulge buildup in the center in the VW galaxies. For either CW or VW, there is no dependency of B/T on $z_{\rm f}$.

We have checked whether the B/T mass ratio is inversely correlated with the time from the last major merger before $z_{\rm f}$. The reasoning behind this test is as following. A major merger largely destroys the galactic disk, but leaves the bulge intact. Hence one should expect that the longer is the time since the last major merger, the more time the disk has to regrow. Therefore, B/T should decrease, if this assumption is valid. However, we did not find any such trend, i.e., B/T appears to be independent of the last merger time.  Of course, this result can be contaminated by a large number of smaller mergers and extensive cold accretion flows, but these are not expected to affect the bulge, and it is doubtful they can damage the disk substantially. Note, that we assume that dry mergers might be absent at these high redshifts. It may be interesting to check this effect at low redshifts, e.g., $z < 1$, where the major mergers become rare and the cold accretion streams run out of steam.   

In section\,\ref{sec:discussion}, we address the bulge kinematics, i.e., whether bulges obtained by stellar surface density decomposition are rotationally or dispersion velocity-supported, by calculating their $v/\sigma$.

\begin{figure*}
\center 
	\includegraphics[width=0.45\textwidth]{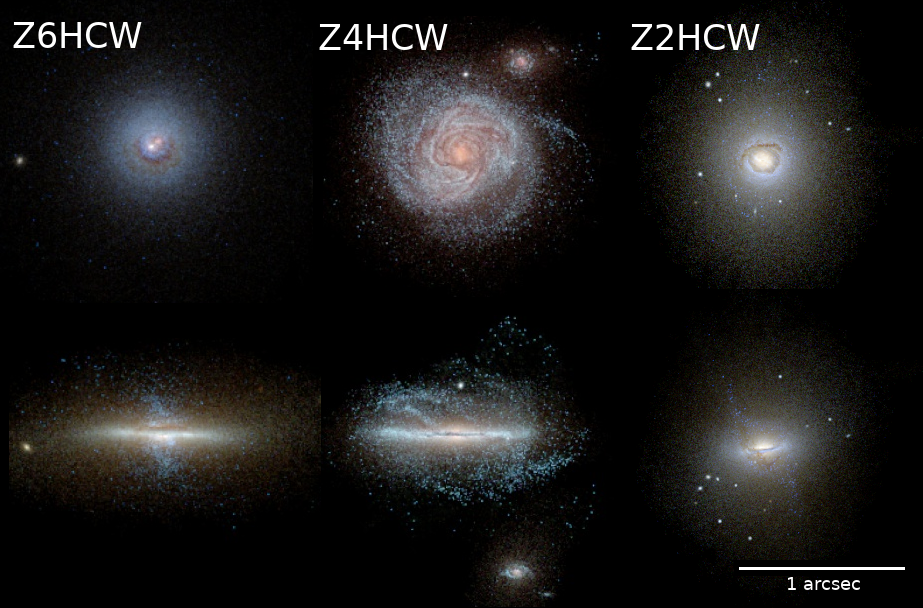}
		\includegraphics[width=0.45\textwidth]{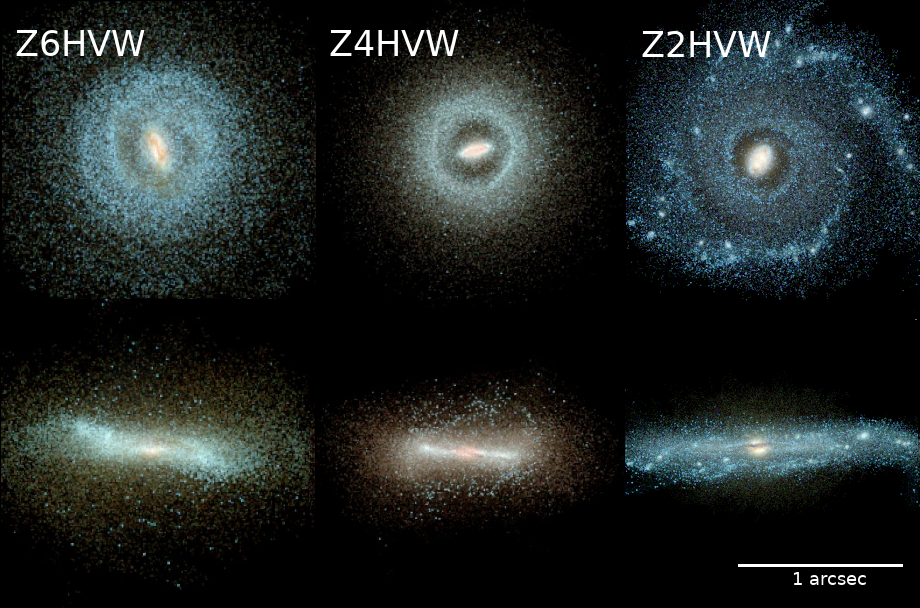}
    \caption{Representative galaxy images at their respective $z_{\rm f}$ and $\delta\sim 3$ with the CW feedback (left frame) and VW (right frame), face-on and edge-on. The images have been obtained by post-processing the stellar light with a 3-D radiation transfer code, and redshifted but not pixelized, nor convolved with the JWST PSF. The dust absorption  has been implemented as well. The three-color mosaic of the JWST filters F070W, F115W and F365W have been used for $z_{\rm f}=6$ and 4 galaxies, and F070W, F115W and F200W for the $z_{\rm f}=2$ galaxies.  
    }
    \label{fig:photo_images}
    \end{figure*} 

\begin{figure*}
\center 
	\includegraphics[width=0.45\textwidth]{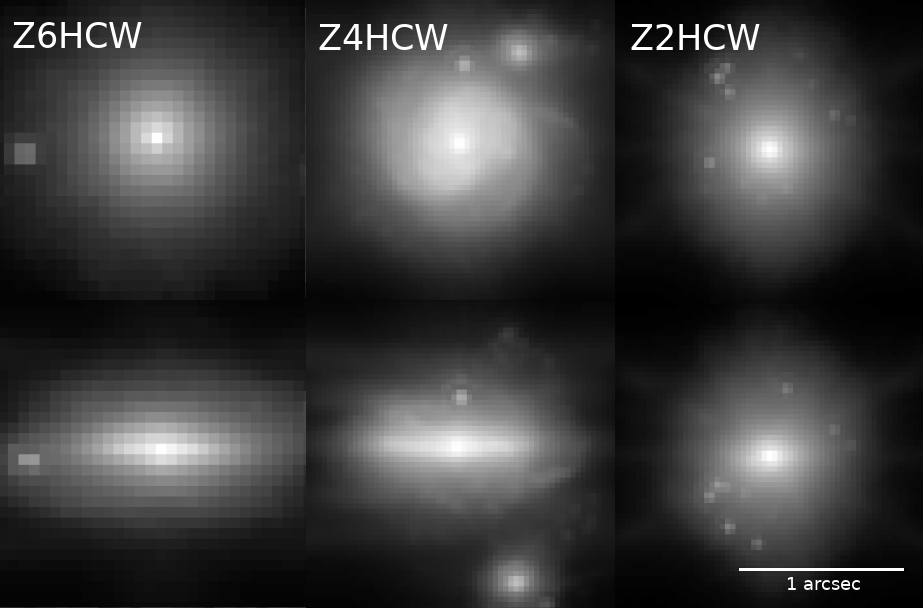}
		\includegraphics[width=0.45\textwidth]{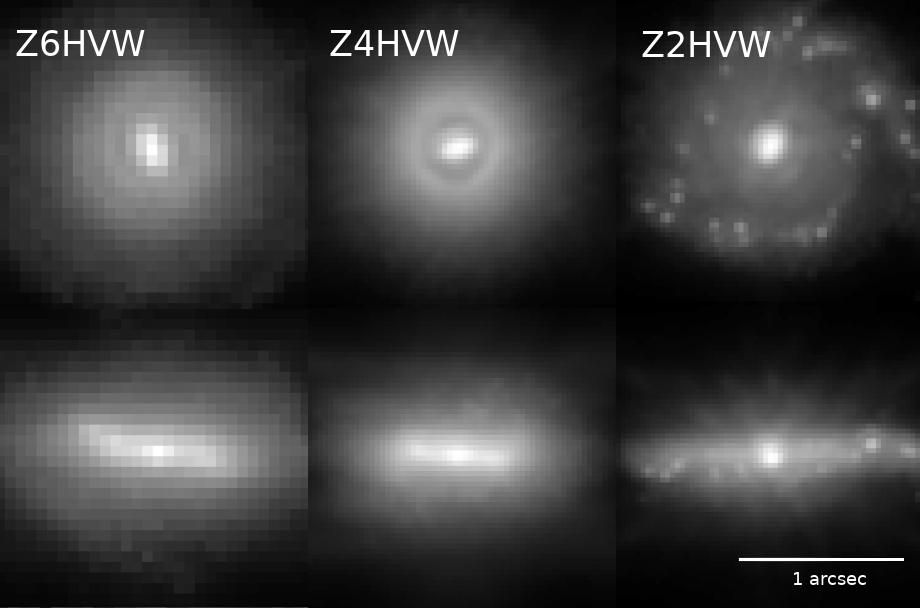}
    \caption{Representative galaxy images at their respective $z_{\rm f}$ and $\delta\sim 3$ with the CW feedback (left frame) and VW (right frame), face-on and edge-on. The images have been obtained by post-processing the stellar light with a 3-D radiation transfer, then redshifted, pixelized, and convolved with the JWST PSF. The dust absorption has been implemented as well. The JWST filter F365W have been used for $z_{\rm f}=6$ and 4 galaxies, and the JWST filter F200W has been used for the $z_{\rm f}=2$ galaxies.}
    \label{fig:photo_images2}
    \end{figure*}

%%%%%%%%%%%%%%%%%%%%%%%%%%%%%%%%%%%%%%%%%%%%%%%%%%%%%%%
\subsection{Disk-bulge decomposition: surface photometry}
\label{sec:photo}
%%%%%%%%%%%%%%%%%%%%%%%%%%%%%%%%%%%%%%%%%%%%%%%%%%%%%%%

 \begin{table}
 \resizebox{\columnwidth}{!}{%
%\caption{  Simulation suite with different galactic wind feedback, constant wind (CW) and variable wind %(VW).}.  
\vspace*{-.1cm}
%\title{  Preliminary Sample Estimates}
\centering
%\resizebox{\columnwidth}{!}{%
\begin{tabular}{ccccccc}
\hline
 $\mathrm{z_{\rm f}}$ & Model Name & $n_{\rm bulge}$ & $R_{\rm e}{\rm (bulge)}\,$kpc &  $R_{\rm e}{\rm (disk)}\,$kpc & Feedback \\
\hline
\hline
    6     & Z6HCW & 1.08 & 0.78 & 2.86 & CW \\
    6     & Z6HVW & 3.70 & 0.49 & 2.80 & VW \\
\hline   
    6     & Z6LCW & 1.59 & 0.55 & 3.75 & CW \\   
    6     & Z6LVW & 0.31 & 0.55 & 2.37 & VW \\   
\hline  
\hline   
    4     & Z4HCW & N/A  & N/A  & 2.37 & CW \\   
    4     & Z4HVW & 0.32 & 0.40 & 2.61 & VW \\  
\hline
    4     & Z4LCW & 1.80 & 0.55 & 2.61 & CW \\
    4     & Z4LVW & 0.81 & 0.58 & 2.76 & VW \\
\hline 
\hline   
    2     & Z2HCW & 0.69 & 0.77 & 2.85 & CW \\   
    2     & Z2HVW & 0.90 & 0.71 & 7.01 & VW \\   
\hline
    2     & Z2LCW & 0.82 & 0.60 & 3.24 & CW \\
    2     & Z2LVW & 0.39 & 0.56 & 2.35 & VW \\

\hline   
\hline
\end{tabular}
}
\caption{Disk-bulge decomposition of modeled galaxies based on the surface photometry at $z_{\rm f}$ in comoving coordinates, as shown and detailed in Figure\,\ref{fig:photo_images}. All models have been fitted with a bulge and an exponential disk components. From left to right: the final redhsift $z_{\rm f}$; the model name; $n_{\rm bulge}$ --- the Sersic index of the bulge component; $R_{\rm e}{\rm (bulge)}$ --- the  half-light radius of the bulge component; $R_{\rm e}{\rm (disk)}$ --- the  half-light radius of the disk component; CW and VW --- the type of the galactic wind feedback.} 
\label{tab:deco_photo}
\end{table}

We have repeated the bulge-disk decomposition for the photometric images of our face-on galaxies. Six synthetic images at $z_{\rm f}=6$, 4, and 2, of CW and VW representative models in the $\delta\sim 3$ overdense regions are shown in Figure\,\ref{fig:photo_images}, face-on and edge-on, by applying a mosaic of three JWST filters for each image, as detailed in section\,\ref{sec:light}. The choice of the filters has been made based on the SED obtained by using \citet{bruz93}. The images have been obtained by post-processing the galaxies with the 3-D radiation transfer (including dust absorption), then redshifted, but not pixelized, nor convolved with the PSF. This figure is shown for comparison only. All galaxies harbor a stellar bar. Because we analyze the properties and evolution of these bars in paper\,II, we only make few general comments here about them and their effect on the galactic morphology. 

We also present the same images but pixelized and convolved with the PSF (Fig.\,\ref{fig:photo_images2}). These degraded resolution images have been used for the decomposition. The JWST filter F365W have been used for $z_{\rm f}=6$ and 4 galaxies, and the JWST filter F200W has been used for the $z_{\rm f}=2$ galaxies. Our results are summarized in Table\,\ref{tab:deco_photo} and the decomposition is shown in the Appendix Figures\,\ref{fig:dec_photoCW} and \ref{fig:dec_photoVW}. In comoving coordinates, the image sizes of simulated galaxies are very similar. Yet the underlying morphology appears to be very different. All galaxies harbor a stellar bar.  

\begin{figure*}
\center 
\includegraphics[width=0.8\textwidth]{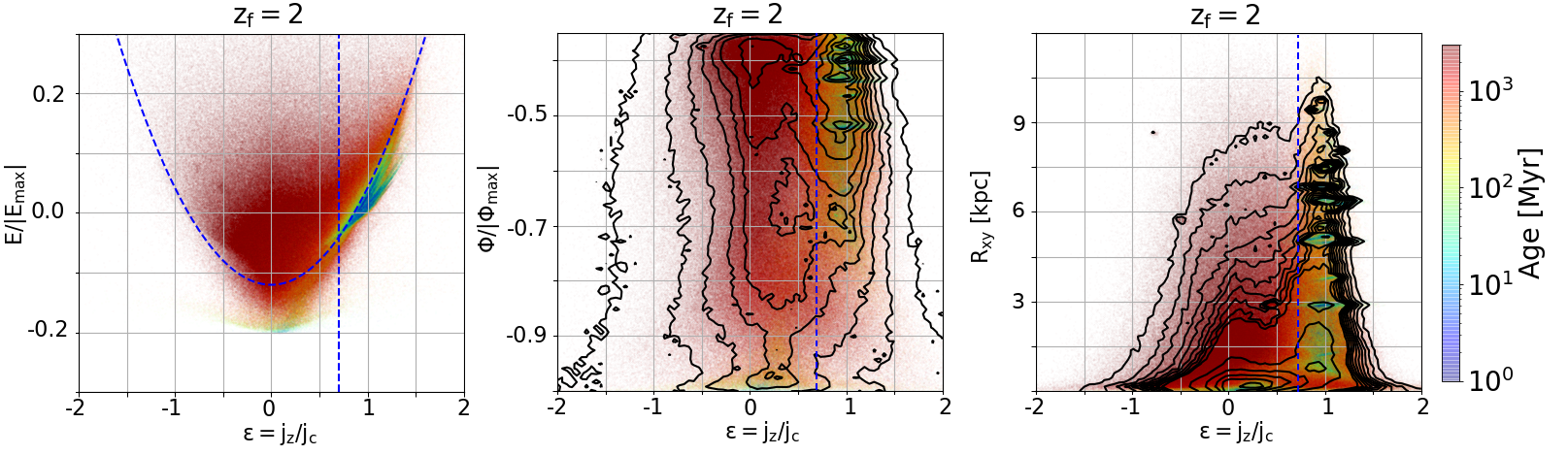}
    \caption{Bulge-disk kinematic decomposition for the stellar Z2LCW model at $z_{\rm f}=2$. The colors correspond to stellar ages. {\it Left:} using total energy, $E/|E_{\rm max}|$, normalized by the maximal energy, versus specific angular momentum normalized by the circular angular momentum, $\epsilon=j_{\rm z}/j_{\rm c}$. The vertical blue dashed line corresponds to $\epsilon=0.7$, and the parabola fitting is described in the text. The disk lies to the right of $\epsilon=0.7$, the bulge --- to the left of $\epsilon=0.7$ and below the parabola, while the stellar spheroid --- to the left of $\epsilon=0.7$ and above the parabola. {\it Middle:} same but using the gravitational potential, $\Phi$ instead of $E$. Contours define the number density of stars on $\Phi-\epsilon$ plane. {\it Right:} Cylindrical radius, $R$, in the $xy$-plane in the comoving frame versus $\epsilon$. Contours define the number density of stars on the $R-\epsilon$ plane. More details in Figure\,\ref{fig:dec_kinem2} and in the text.
    }
    \label{fig:dec_kinem1}
    \end{figure*} 

In the degraded resolution Figure\,\ref{fig:photo_images2}, the CW galaxies at $z_{\rm f}=6$ and 4 include the disk component. But the $z_{\rm f}=2$ Z2HCW galaxy, while still possessing the disk component, has developed an equally important spheroid. For the CW galaxies, the bars are not visible, but for the VW objects they are still recognisable. At the final redshift, only $z_{\rm f}=4$ Z4HCW and $z_{\rm f}=2$ Z2HVW galaxies exhibit traces of spiral structure, yet the intermittent and well-developed spiral structure has been observed during the evolution towards $z_{\rm f}$ in almost all objects. We note that all the CW galaxies possess small (compared to the disk size) stellar bars, as can be observed in high-resolution images of Figure\,\ref{fig:photo_images} and analyzed in Paper\,II, but some of these bars appear not to drive the spiral structure.

The VW galaxies' morphology differs profoundly from that of the CW ones. Again, all of the VW galaxies host stellar bars, still recognisable in the degrade figure, because the size of these bars is substantially larger than in CW ones. Each of these bars appears to drive a pair of spiral arms, which wind up tightly towards $z_{\rm f}$, but under the degraded resolution they are washed out, except in Z2HVW. These bars determine the sites of star formation. Outside the bar region, the disks are depopulated from stars, which delineates the corotation region\footnote{The corotation is defined as the radius where the bar pattern speed is equal to the circular stellar frequency} (CR). Outside the CR, the spirals arms form a frequently observed outer rings, situated close to the outer Lindblad resonance (OLR). The stars amplify this difference due to the ongoing star formation in the spiral arms which form the rings. We detect the associated spirals arms in gas, and provide the ALMA images (Fig.\,\ref{fig:ALMA}).  

For the photometric decomposition, we find that $n_{\rm b} < 1$ for the VW galaxies, except Z6HVW at $z_{\rm f}=6$. We do not find substantial differences in $n_{\rm b}$ indexes between the CW and VW galaxies. We also find that this index is rather independent of $z_{\rm f}$, except at $z_{\rm f}=6$, where two galaxies have substantially higher $n_{\rm b}$. B/T are mixed for CW and VW objects, and are rather independent of $z_{\rm f}$, i.e., flat.  

To compare $n_{\rm b}$ for stellar surface density and photometric decomposition, we find that for the CW galaxies, $n_{\rm b}{\rm (photometric)} < n_{\rm b}$(stellar). For the VW galaxies, $n_{\rm b}{\rm (photometric)} > n_{\rm b}$(stellar). On the other hand, for the B/T mass ratio, for both the CW and VW galaxies, B/T(stellar) $\sim $ B/T(photometric). 

Comparing with the de Vaucouleurs law of $n_{\rm b}=4$ \citep{vau48,binn08}, we note that most of the photometric indexes are small, i.e., $n_{\rm b}\ltorder 1$, with the exception of Z6HVW which has  $n_{\rm b}\sim 3.7$. Two CW galaxies have  $n_{\rm b}\sim 1.59$ and 1.8, and another galaxy appears bulgeless, Z4HCW. Galaxies, especially the VW ones, have $n_{\rm b} < 1$ and harbor stellar bars which have a flat light distribution.  

%%%%%%%%%%%%%%%%%%%%%%%%%%%%%%%%%%%%%%%%%%%%%%%%%%
\subsection{Disk-bulge decomposition: kinematics}
\label{sec:kinematics}
%%%%%%%%%%%%%%%%%%%%%%%%%%%%%%%%%%%%%%%%%%%%%%%%%%
 
 \begin{figure*}
\center 
\includegraphics[width=0.8\textwidth]{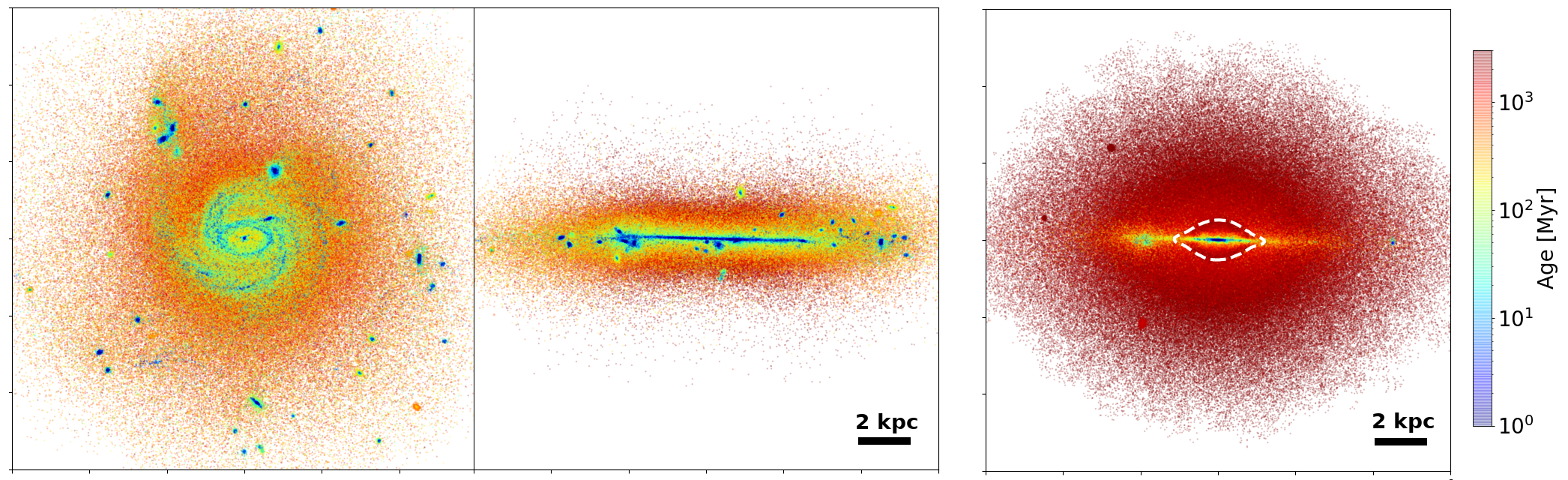}
    \caption{Bulge-disk kinematic decomposition of the Z2LCW model at $z_{\rm f}=2$ (additional details in Fig.\,\ref{fig:dec_kinem1} and in the text). The colors correspond to stellar ages. Shown from left to right are the face-on disk component, the edge-on disk component, and the edge-on stellar spheroid. The white dashed line shows the outline of the stellar bulge.}
    \label{fig:dec_kinem2}
    \end{figure*}  
 
 \begin{figure}
\center 
	\includegraphics[width=0.49\textwidth]{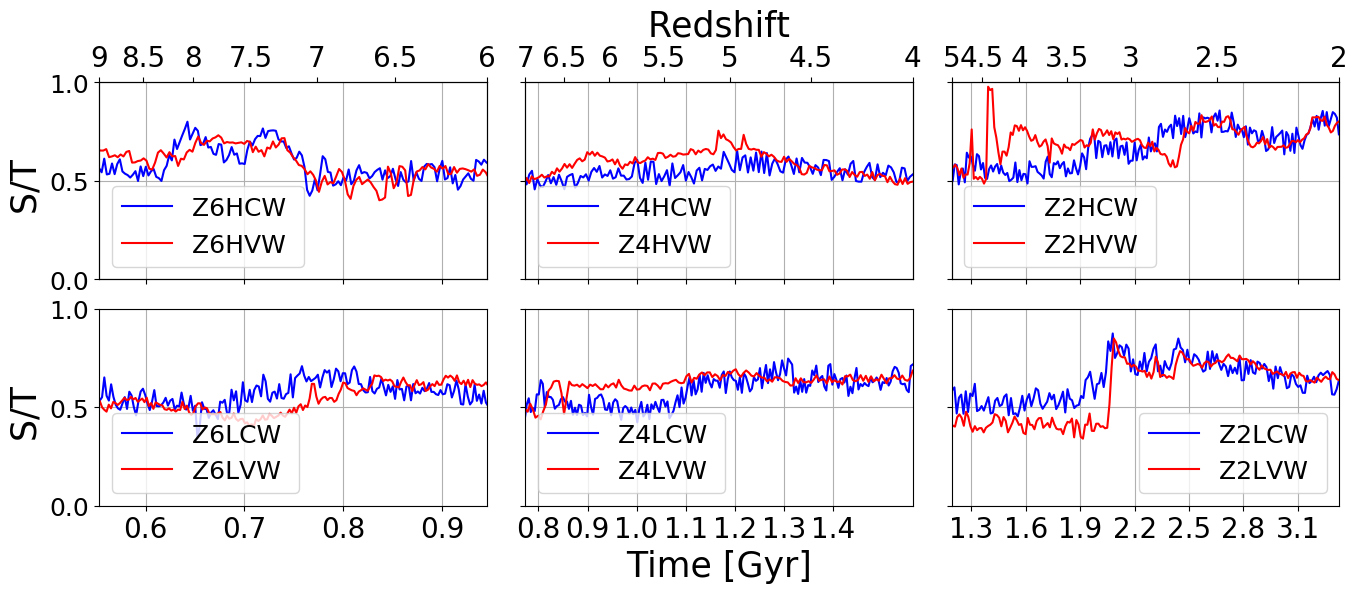}
    \caption{Evolution of the stellar spheroid-to-total stellar mass ratio, S/T, for all galaxy models with the final redshifts of $z_{\rm f}=6, 4,$ and 2, for CW (blue lines) and VW (red lines).}
    \label{fig:ST}
    \end{figure}  

The kinematic bulge-disk decomposition is the next step. The Figure\,\ref{fig:dec_kinem1} (left) displays the decomposition in the total specific energy normalized by the maximal energy, $E/|E_{\rm max}|$, versus normalized specific angular momentum, $\epsilon = j_{\rm z}/j_{\rm c}$, where $j_{\rm c}$ is the circular angular momentum --- $j_{\rm z}$ and $j_{\rm c}$ are calculated for each stellar particle. One can distinguish three main regions on the $E/|E_{\rm max}| - \epsilon$ plane: starting with the stellar disk for $\epsilon > 0.7$, dominated by rotation. To the left of the vertical dashed line lies the bulge below the dashed parabola (see below for details), and the stellar spheroid (above the parabola). The parabola choice comes from the dependence of energy on $j^2$, and roughly separates the bulge from the stellar spheroid. 

We fit the parabola by binning $\epsilon$ and finding the minimal gradient along the $E/|E_{\rm max}|$ axis, for every bin with $\epsilon \ltorder 0$. The final parabola shape has been obtained by applying a least square fit for the highest (point) density contour in this phase diagram. The negative $\epsilon$ is preferred because it is not contaminated by the disk, and by assuming that the parabola is symmetric with respect to $\epsilon=0$ axis. As an example, we perform the decomposition for the Z2LCW galaxy at $z_{\rm f} = 2$. 

The total stellar mass of this galaxy is $M_*{\rm (galaxy)} = 5.33\times 10^{10}\,M_\odot$. Within the bulge radius, we add the stellar spheroid to the bulge, as it is difficult to distinguish between them kinematically. This results in the mass ratio B/T\,$= M_*({\rm bulge})/M_*({\rm galaxy})\sim 0.32$, which is very close with our surface density decomposition where for this galaxy we have obtained B/T\,$\sim 0.38$ (section\,\ref{sec:density}). Results of this decomposition in the real space can be seen in Figure\,\ref{fig:dec_kinem2}. The disk and bulge are well separated, while the spheroid has a residual disk component, which is not significant.  

{  The kinematic decomposition disk-spheroid has been attempted in numerous publications before \citep[e.g.,][]{abadi03a,abadi03b,croft09,scanna09,mari14,rosa20}. Mostly, they used separation based on $\epsilon$, which can be traced back to the foundation of classical mechanics \citep[e.g.,][]{lan60}. Typically, stars with $\epsilon < 0.7-0.8$ have been attributed to the spheroidal component. Moreover, it was assumed that the spheroid has a spherical shape and no angular momentum, so its $\epsilon$ distribution is symmetric with respect to zero. However, spheroids can have an arbitrary shape and include bulges which can be rotationally supported \citep[e.g.,][]{kor04}. Moreover, some fraction of stars in galactic disks and stellar bars can be on retrograde orbits \citep[e.g.,][]{coll18}. }

{  A different method for the kinematic decomposition when distribution of angular momentum versus projected distance in the disk plane was applied as well \citep{scanna09}. In many cases this resulted in well separated populations, disks and spheroids, when stellar orbits have been restricted to small inclination with respect to the disk plane. But it does not separate the bulges from extended spheroids. Moreover, the method is time consuming, requires individual inspection, and actually has been replaced by a simple $\epsilon=0.5$ separation by the authors, which displayed the evolution of disk-to-total mass ratio for $z < 3$. }

For brevity, we have tried two additional planes for decomposition in Figure\,\ref{fig:dec_kinem1} (middle and right frames, respectively), by replacing the normalized total energy with the normalized gravitational potential $\Phi/|\Phi_{\rm max}|$, and, finally, replacing it by the cylindrical radius, using the $R - \epsilon$ plane, in comoving coordinates. The colors are the stellar ages in all three decompositions, and the contours represent the density of the stellar particles in both planes. The final results of all kinematic bulge-disk decompositions do not differ significantly from each other.

{  Therefore, our kinematic decomposition goes one step ahead by separating the bulge from the surrounding spheroidal component. This has allowed us to recalculate the rotational support for the stellar disks in our models, with bulges and spheroids excluded. Results are given in the last column of Table\,\ref{tab:BTcompare}. Averaging the numbers for each $z_{\rm f}$, we observe that the rotational support increases with time, i.e., it is $<v_{\rm los}/\sigma_{\rm z}>\sim 3.7$ for $z_{\rm f}=6$, 3.8 for $z_{\rm f}=4$, and 5.3 for $z_{\rm f}=2$. This numbers exceed substantially the numbers displayed in Figure\,\ref{fig:vsigma_stars}, where we used the total stellar surface density (i.e., using the Sersic method), which included the contributions from the bulge and the spheroid. Moreover, using the kinematic decomposition, we do not find any significant difference between the stellar disk rotational support in objects residing in high versus low overdensity. But we do record a significantly larger rotational support in VW models compared to CW ones, especially for $z < 6$. }

Table\,\ref{tab:BTcompare} shows the results of the three decomposition methods (i.e., the stellar surface density, the surface photometry, and kinematic) of the B/T mass ratio, and the stellar spheroid-to-total stellar mass ratio, S/T (Fig.\,\ref{fig:ST}). Here, we consider spheroid to have $\epsilon \ltorder 0.7$, i.e., we include the bulge in the spheroid. The range of S/T appears to be quite narrow, $0.5-0.8$, and not much difference between the CW and VW galaxies is observed. The S/T appears flat for $z_{\rm f=6}$ and 4 galaxies, except some variability. A tendency to rise S/T is exhibited in $z_{\rm f}=2$ galaxies, and can be traced to multiple mergers and other perturbers, as these galaxies are analyzed between $z=5$ to 2 --- the longest time period of all the models addressed here and therefore the most affected by internal and external factors. We discuss these in section\,\ref{sec:discussion}.

 \begin{table}
 \resizebox{\columnwidth}{!}{%
\vspace*{-.1cm}
%\title{  Preliminary Sample Estimates}
\centering
%\resizebox{\columnwidth}{!}{%
\begin{tabular}{cccccccc}
\hline
 $\mathrm{z_{\rm f}}$ & Model Name & B/T     & B/T         & B/T       & S/T &   $v_{\rm los}/\sigma_{\rm z}$  \\
                      &           & stellar & photometric & kinematic & kinematic & kinematic \\
\hline
\hline
    6     & Z6HCW & 0.52 & 0.29 & 0.36 & 0.57 & 3.8 \\
    6     & Z6HVW & 0.53 & 0.17 & 0.37 & 0.53 & 3.8 \\
\hline   
    6     & Z6LCW & 0.35 & 0.10 & 0.30 & 0.57 & 4.0 \\   
    6     & Z6LVW & 0.48 & 0.14 & 0.46 & 0.63 & 3.4 \\   
\hline  
\hline   
    4     & Z4HCW & 0.29 & 0.00 & 0.38 & 0.59 & 3.8 \\   
    4     & Z4HVW & 0.50 & 0.09 & 0.39 & 0.50 & 3.9 \\  
\hline
    4     & Z4LCW & 0.59 & 0.58 & 0.58 & 0.75 & 2.9 \\
    4     & Z4LVW & 0.74 & 0.50 & 0.62 & 0.67 & 4.6 \\
\hline 
\hline   
    2     & Z2HCW & 0.49 & 0.23 & 0.43 & 0.80 & 2.5 \\   
    2     & Z2HVW & 0.81 & 0.27 & 0.54 & 0.80 & 8.3 \\   
\hline
    2     & Z2LCW & 0.38 & 0.11 & 0.32 & 0.62 & 5.3 \\
    2     & Z2LVW & 0.68 & 0.22 & 0.52 & 0.65 & 5.2 \\

\hline   
\hline
\end{tabular}
}
\caption{Comparison of disk-bulge decomposition of modeled galaxies based on the three methods at $z_{\rm f}$.  From left to right: the final redhsift $z_{\rm f}$; the model; B/T --- bulge-to-total stellar mass, based on stellar surface density, surface photometry, and on kinematics; S/T --- stellar spheroid-to-total stellar mass; stellar disk rotational support: $v_{\rm los}/\sigma_{\rm z}$ for the stellar disk component using kinematic decomposition at $z_{\rm f}$ (i.e., without the bulge and spheroid). } 
\label{tab:BTcompare}
\end{table}  

%%%%%%%%%%%%%%%%%%%%%%%%%%%%%%%%%%%%%%%%%%%%
\subsection{Migration of stellar population}
\label{sec:migration}
%%%%%%%%%%%%%%%%%%%%%%%%%%%%%%%%%%%%%%%%%%%%

\begin{figure}
\center 
	\includegraphics[width=0.45\textwidth]{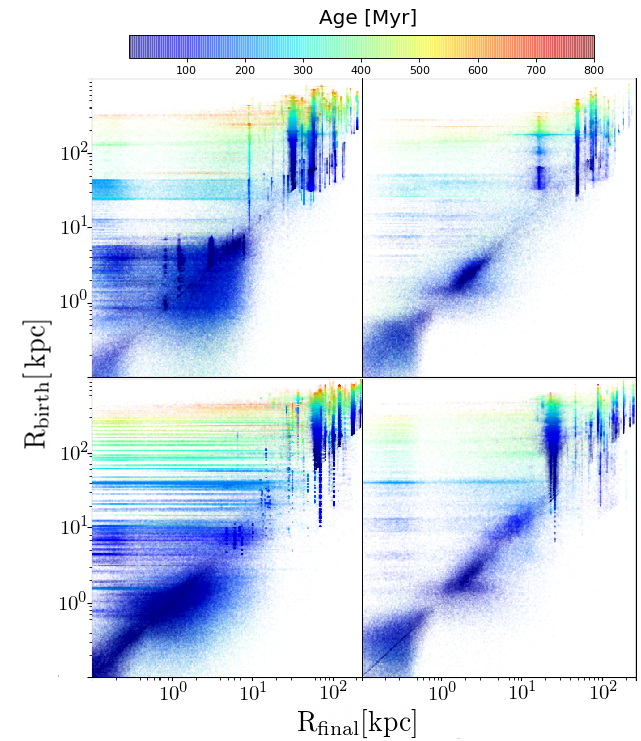}
    \caption{Stellar migration from the birth radius, $R_{\rm birth}$, to the final radius, $R_{\rm final}$, at $z_{\rm f}=6$. Shown are models Z6LCW and Z6LVW (top), and Z6HCW and and Z6HVW (bottom). The vertical lines in the upper right corner, $r \gtorder 20\,h^{-1}$ represent the stars found in substructures that after some time dissolve in the halo and contribute to the halo stellar population. All in comoving coordinates. Colors represent stellar ages given in Myr. More details in the text.}
    \label{fig:migration1}
    \end{figure} 

\begin{figure}
\center 
	\includegraphics[width=0.45\textwidth]{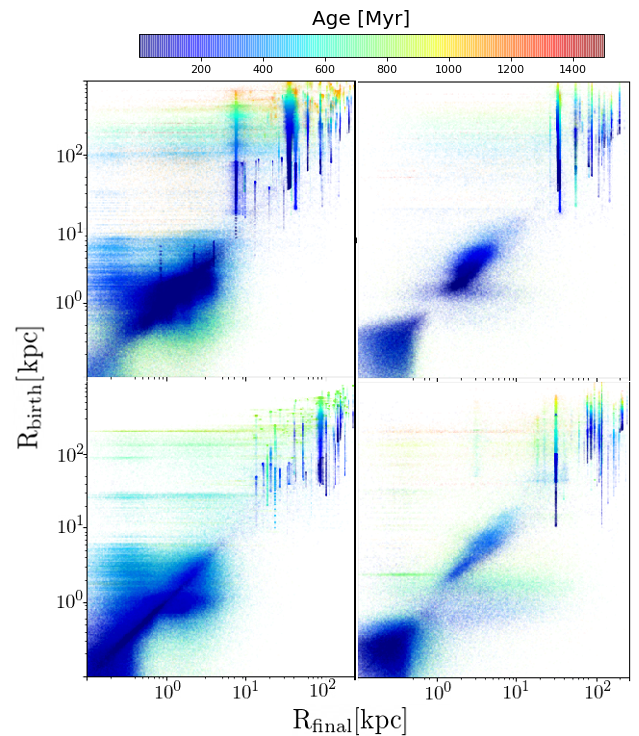}
    \caption{Stellar migration, as in Figure\,\ref{fig:migration1} but for $z_{\rm f}=4$, and for models Z4HCW and Z4HVW (top), Z4LCW and  Z4LVW (bottom).}
    \label{fig:migration2}
    \end{figure} 

 \begin{figure}
\center 
	\includegraphics[width=0.45\textwidth]{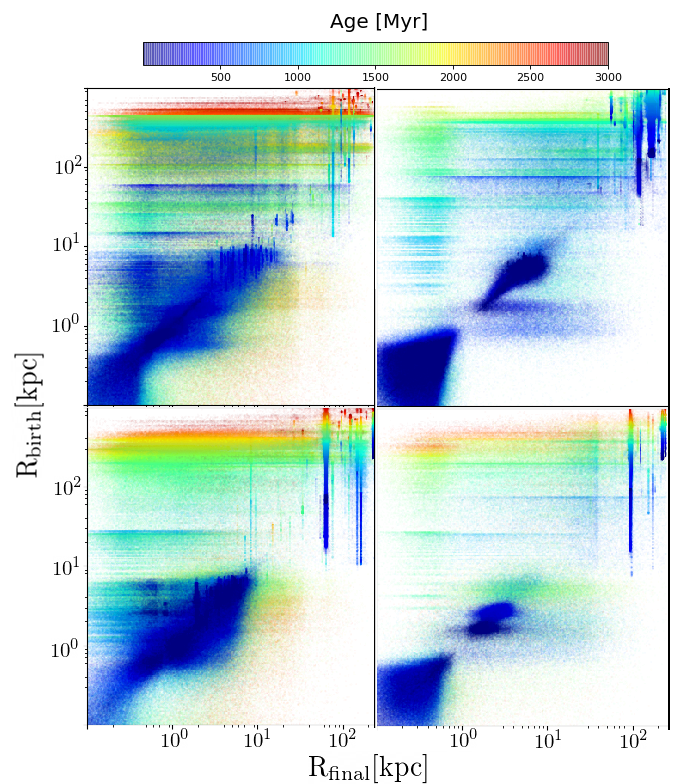}
    \caption{Stellar migration, as in Figure\,\ref{fig:migration1} but for $z_{\rm f}=2$, and for models Z2HCW and Z2HVW (top), and Z2LCW and Z2LVW (bottom).}
    \label{fig:migration3}
    \end{figure} 

\begin{figure}
\center 
	\includegraphics[width=0.45\textwidth]{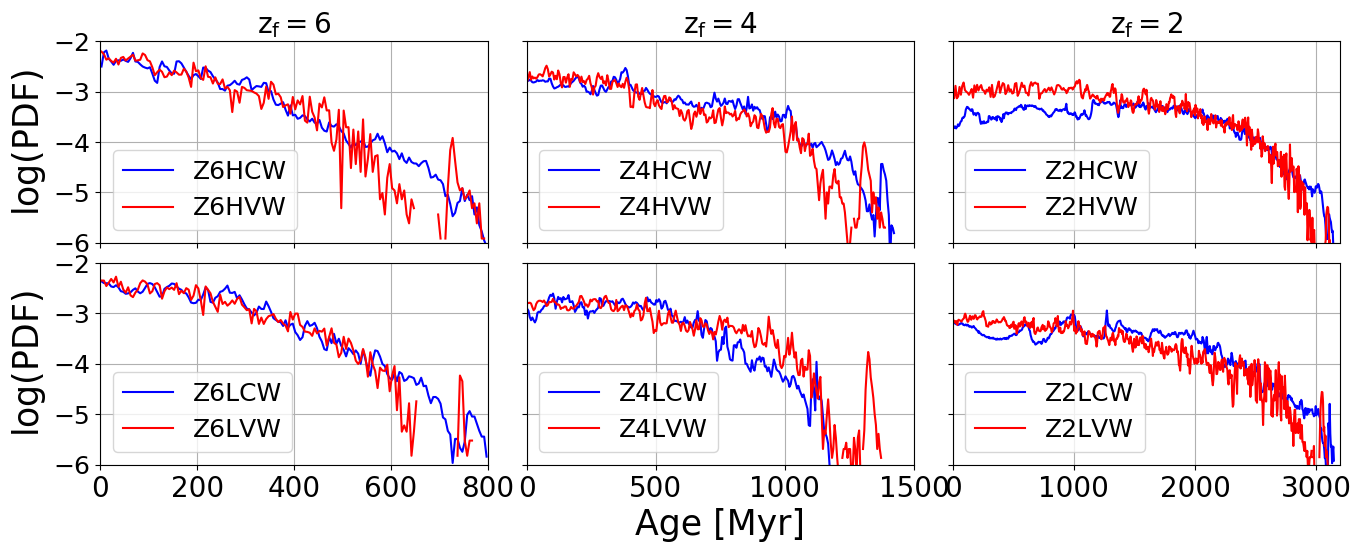}
    \caption{The PDF of stellar ages in simulated galaxies. The time shown is the lookback time from $z_{\rm f}$ for each model. The curves have been normalized by the total number of stars in the parent galaxies at $z_{\rm f}$ for CW and VW separately. The ages were binned in 5\,Myr. The VW models are given by the red lines, and CW models by the blue lines.}
    \label{fig:PDFage}
    \end{figure} 

Figures\,\ref{fig:migration1}, \ref{fig:migration2}, and \ref{fig:migration3} show the migration of the stellar population between the stellar birth position, $R_{\rm birth}$, and the final stellar position, $R_{\rm final}$ --- an approach first used by \citet{yu20}. All in comoving coordinates, The color palette exhibits the stellar ages. We do not subtract the contributions from subhalos, partly because they are well separated anyway and located in the upper right corners, and partly because the stellar subhalos will ultimately contribute to the formation of the outer stellar halos. 

The stellar ages correlate with $R_{\rm birth}$, decreasing inwards, i.e., from larger to smaller radii. In fact, the stellar ages are substantially larger for the stellar population in the parent halos  compared to galaxies --- a clear sign that the outer halo stars originate outside the main galaxy. For example, in Figure\,\ref{fig:migration1}, for CW and $z_{\rm f}=6$, the outer halo stars have ages of $\sim 500-700$\,Myr, while for the innermost halo of $< 75$\,kpc (comoving coordinates) their ages are less than 200\,Myr.

The approximate sizes of embedded galaxies are clearly delineated by the young (blue color) stellar populations --- all lying on the diagonal $R_{\rm birth}-R_{\rm final}$ lines. The "puffing" of this component results from the combined action of disk rotation and stellar dispersion velocities which contribute to the stellar spheroid. 

We can divide this evolution very roughly into three regions, based on the stellar ages and associated stellar densities. The outer region includes the host halo and extends  from $R_{\rm vir}$ towards $45-75\,{\rm kpc}$, in comoving coordinates. This inner boundary, besides displaying an increase in the stellar density, is accompanied by a sharp difference in the stellar ages. The intermediate region (range) lies between $45-75\,{\rm kpc} \rightarrow 10-16\,{\rm kpc}$, and the innermost region is found inside $10-16\,{\rm kpc}$. The latest division is based on additional increase in the stellar density with decreasing $R_{\rm birth}$. These divisions can be applied to both CW and VW models at all $z_{\rm f}$.

The two outer regions characterize the stellar population in the host halos, while the inner region belongs to the growing galaxies. We find that for two outer regions in the CW models, the typical $R_{\rm final}\ltorder R_{\rm birth}$. Hence the stars migrate inwards in the DM halos, which can be observed in the form of horizontal bands. But for the inner region, $R_{\rm final}\gtorder R_{\rm birth}$, so the stars migrate outwards, which again is associated with horizontal bands outside the diagonal. 

For the VW models, the stellar population of parent halos is visibly smaller, and in the galaxy region a number of stars have $R_{\rm final}\gtorder R_{\rm birth}$. Overall, more stars remain in their original positions in VW than in CW models and do not migrate at $z_{\rm f}=6$. This of course can be a temporary effect, as the galaxies are very young, even for $z_{\rm f}=2$ the evolutionary timescale is limited by about 1\,Gyr.

In addition to the inward migration of halo stars which increases with decreasing $z_{\rm f}$, we observe the outwards migration from the innermost region corresponding to galaxies. This outward migration is hardly discernible in $z_{\rm f}=6$ galaxies, but appears in $z_{\rm f}=4$ objects and strengthens further in $z_{\rm f}=2$ galaxies. The outward migration is much more pronounced in the VW models --- a reflection of a stronger feedback which generates stronger gas outflows. It seems an inescapable conclusion that these stars have been born within the outflowing gas, as argued in \citet{yu20}. 

The galactic stellar population is represented by the `fat' blobs centered on the diagonal lines in all Figures. Typically, this population is young and can be divided kinematically into two subgroups --- rotationally supported and having nonnegligible dispersion velocities. We have analyzed their kinematics in sections\,\ref{sec:general} and \ref{sec:kinematics}.

An interesting feature can be observed clearly in all VW models in Figures\,\ref{fig:migration1}, \ref{fig:migration2}, and \ref{fig:migration3} at around $1.5\,$kpc from the center in comoving coordinates, on the diagonal line $r_{\rm birth}=r_{\rm final}$. The origin of this feature is real and can be observed also in Figure\,\ref{fig:photo_images} as a dark circle surrounded by a bright circle. These rings are related to the action of stellar bars which can be especially noticeable in the VW galaxies, where these bars are stronger and larger in size compared to the size of the parent galaxy \citep{bi21}. The associated darker region around the corotation radius is partly depopulated by the bars which push the gas and stars out. The bright rings are located at the positions of the outer Lindblad resonances (OLRs) and are frequently observed in barred galaxies at low redshifts.  

We also present the Probability Distribution Functions (PDFs) of stellar ages as a lookback time for each modeled galaxy (Fig.\,\ref{fig:PDFage}). The CW and VW models differ at higher redshifts (with respect to their $z_{\rm f}$), with CW models typically lying above the VW ones. The $z_{\rm f}=6$ models show rising curves towards low ages, reflecting the increasing SFRs at lower redshifts. The $z_{\rm f}=4$ models display a mixture of rising/flat PDF curves at low ages, while the $z_{\rm f}=2$ models even show flat/declining curves there, e.g., Z2HCW. In this object, the SFR declines after $z\sim 3$. Overall, these PDFs are in agreement with the expected SFR peaking around $z\sim 4-2$.

%%%%%%%%%%%%%%%%%%%%%%%%%%%%%%%%%%%%%%%%%%%%
\subsection{Galactic gas morphology}
\label{sec:gas}
%%%%%%%%%%%%%%%%%%%%%%%%%%%%%%%%%%%%%%%%%%%%    

 \begin{figure*}
\center 
	\includegraphics[width=0.45\textwidth]{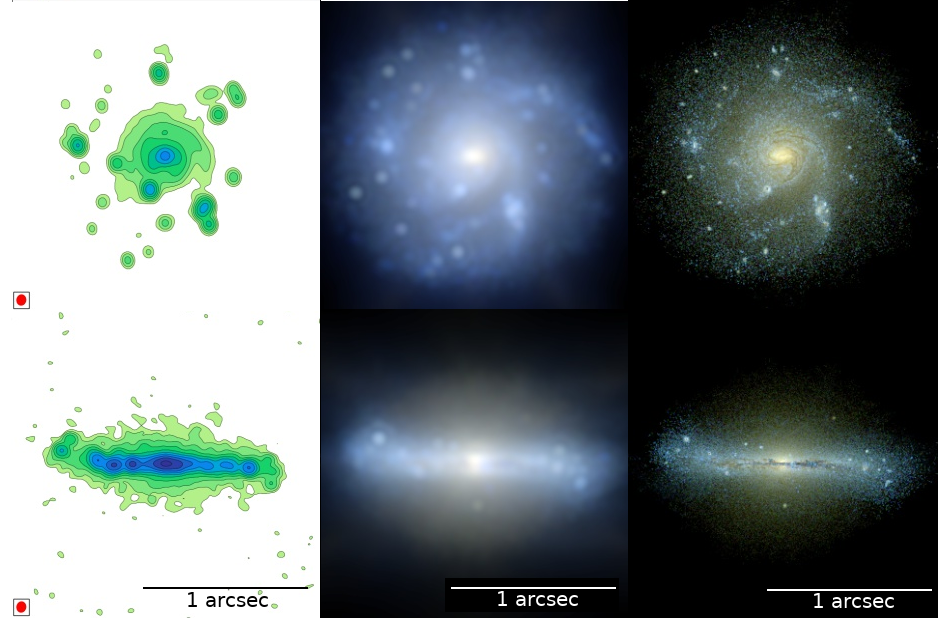}
	\includegraphics[width=0.45\textwidth]{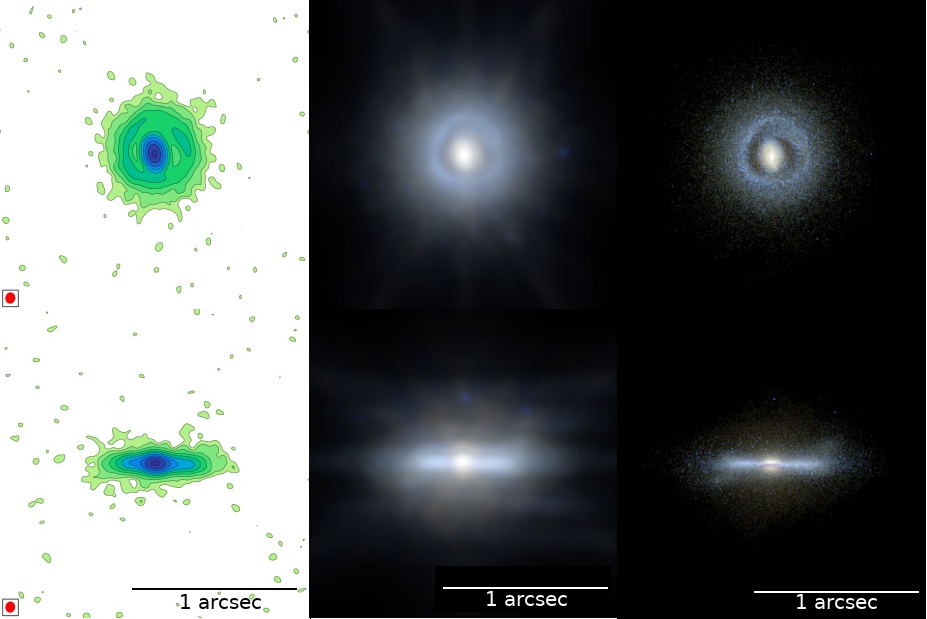}
    \caption{The JWST and ALMA images for Z2LCW (left triad) and Z2LVW (right triad) galaxies at $z=2$ (see section\,\ref{sec:photo} for details). Each image is shown in the face-on and edge-on projections. The details of the JWST images consisting of a single filter and convolved with the PSF image (middle) and the 3-filter mosaic, not convolved with the PSF (right) are given in   Figure\,\ref{fig:photo_images2} and Figure\,\ref{fig:photo_images}, respectively. The ALMA imaging used the maximal bandwidth of 7.5\,GHz centered at 93.7\,GHz (band 3) with an exposure of 3 hours (single visit). The frame extent is similar to that of the JWST image of 1.8", meaning that the galaxy is about 1" in diameter. Resolution for ALMA: major axis = 0.061", minor axis = 0.053" (indicated by red dots). Contours and colors in the ALMA images represent the surface brightness. All contours are plotted at every 2$\sigma$ from 3$\sigma$ to 11$\sigma$ and at every 5$\sigma$ above 11$\sigma$. Comparing the JWST and ALMA images, one can distinguish for the presence of the outer spiral arms and elongated isophotes in the central region --- a signature of the bar in the CW galaxy. The bar in the VW galaxy is much more pronounced and is surrounded by a pair of tightly wound spirals starting at the bar major axis and forming an outer ring at around the outer Lindblad Resonance (OLR).}
    \label{fig:ALMA}
    \end{figure*} 
 
  \begin{figure}
\center 
	\includegraphics[width=0.4\textwidth]{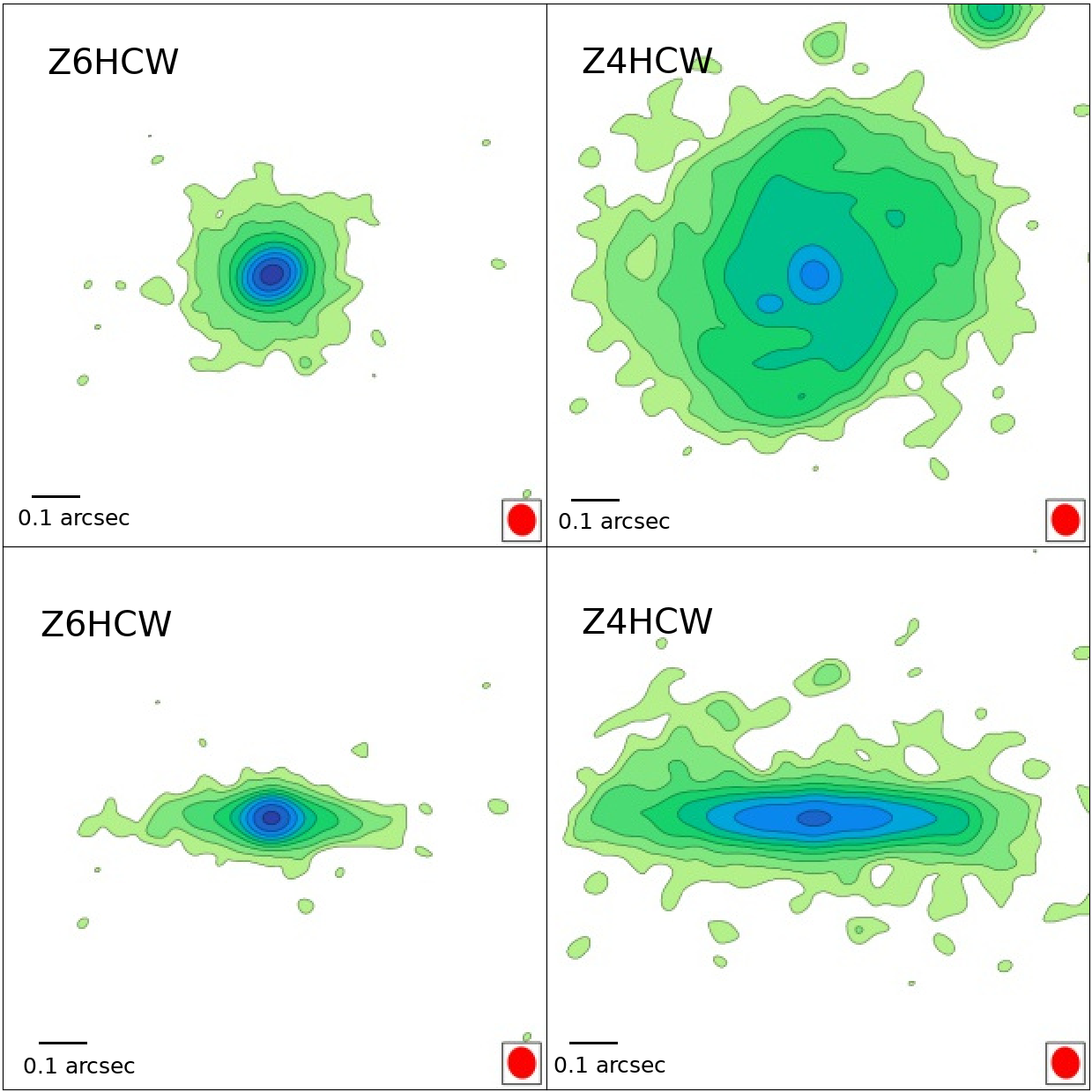}
    \caption{The ALMA images for Z6HCW galaxy at $z=6$ (left) and Z4HCW galaxy at $z=4$. Each image is shown in the face-on and edge-on projections.  The ALMA imaging used the maximal bandwidth of 7.5\,GHz centered at 93.7\,GHz (band 3) with an exposure of 3 hours (single visit). Resolution: major axis = 0.061", minor axis = 0.053" (indicated by red dots). Contours and colors in the ALMA images represent the surface brightness. All contours are plotted at every 2$\sigma$ from 3$\sigma$ to 11$\sigma$ and at every 5$\sigma$ above 11$\sigma$.}
    \label{fig:ALMA2}
    \end{figure} 
 
\begin{figure}
\center 
	\includegraphics[width=0.49\textwidth]{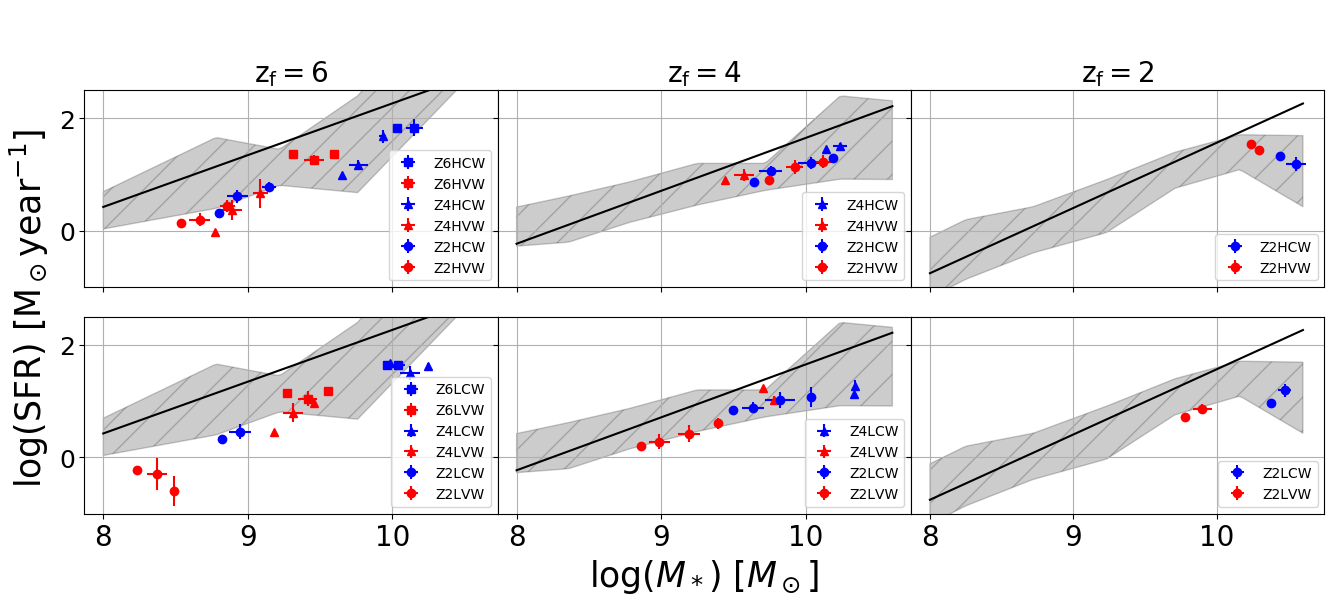}
    \caption{The SF main sequence for simulated galaxies with $z_{\rm f}$: the SFR versus galaxy stellar mass, where we used $M_*(z)$, for CW models (blue) and VW (red). $M_*(z)$ was binned in half dex. The solid lines represent the best fit linear relation for the observational main sequence with the evolving slopes of 0.92 for $z\sim 6-5$, 1.02 for $z\sim 4-3$, and 1.04 for $z\sim 2-1.3$  \citep{santi17}, for $z_{\rm f}=6$, 4, and 2, respectively. The gray colored bands correspond to one $\sigma$ error. More details in the text.}
    \label{fig:SFR_Mstar}
    \end{figure}  
 
\begin{figure}
\center 
	\includegraphics[width=0.48\textwidth]{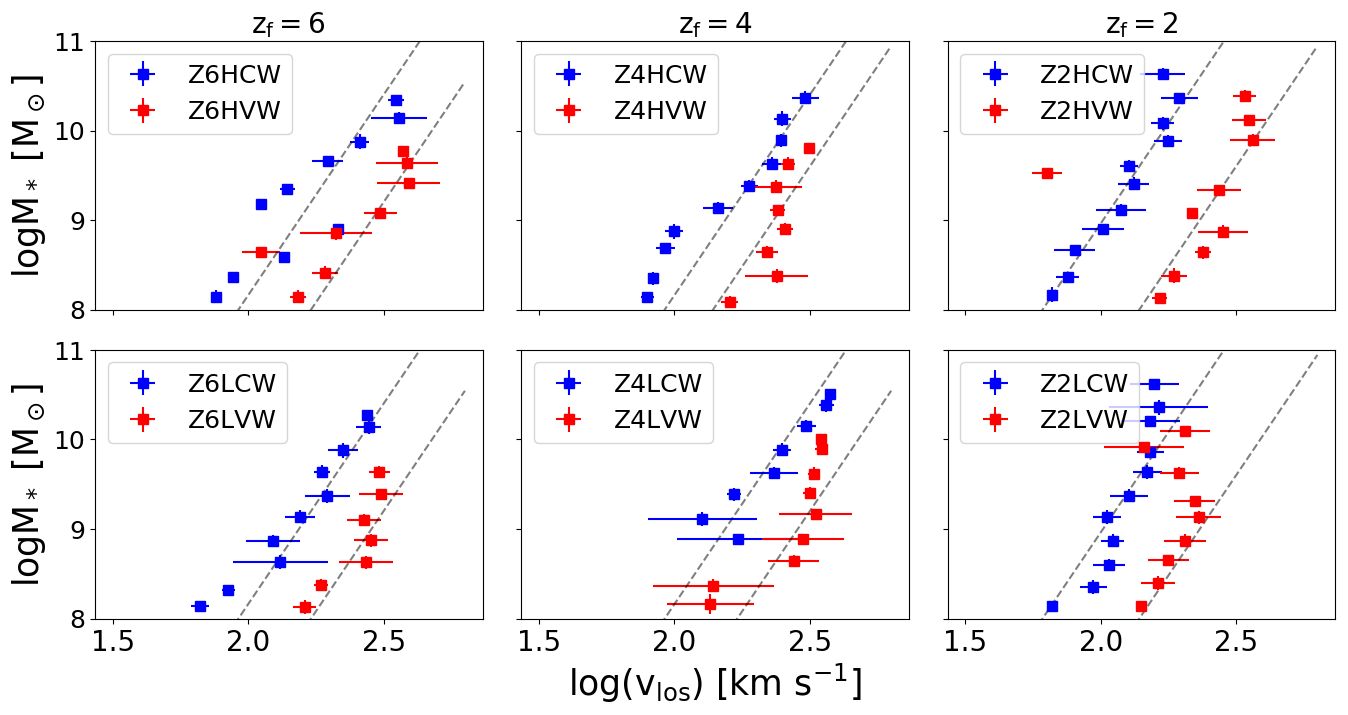}
    \caption{The TF relation for simulated galaxies: the maximal rotational velocity, $v_{\rm max}$, versus $M_*$ for simulated galaxies, for CW models (blue dots) and VW (red dots). Each dot correspond to $M_*(z)$ and $v_{\rm max}(z)$ at a specific redshift binned by the mass. Each dashed line is an average which has the slope of 4.48, obtained for local field galaxies in the NIR by \citet{torr11}. We have adjusted the y-intercept for each model separately.}
    \label{fig:maxvel_Mstar}
    \end{figure}       
 
We have used ALMA to test the gas morphology in modeled galaxies at $z = 2$, 4 and 6 and compared the gas (i.e., its dust component in emission detected by ALMA) and the stellar morphologies detected with the JWST, the latter only for $z=2$. We ask whether morphologies detected by ALMA and the JWST are compatible in large and small-scale details. Using the Z2LCW galaxy in the left frame of Figure\,\ref{fig:ALMA}, the gaseous disk is clearly detected by both instruments. ALMA confirms the warping of this disk, as seen by the JWST. The stellar bar, which is horizontal on the JWST image, is observed also with ALMA. We can detect the presence of two spiral arms originating on the bar major axis. These arms are not only delineated by the brightness contours but also by enhanced emission. A population of starburst clumps in the outer disk shows counterparts both in ALMA and the JWST, confirming the gas presence there, which fuels the ongoing star formation. 
    
The right pair of images in Figure\,\ref{fig:ALMA} correspond to Z2LVW galaxy which shows a profound difference with its CW counterpart. Again, the disk is detected by both instruments, being slightly smaller in extent. On the other hand, its morphology is dominated by the central bar which drives a pair of prominent spirals. These form a stellar ring around the position of the OLR. The gaseous disk extends passed the OLR. In a stark contrast between the VW and CW models, the number of star forming clumps is negligible, both in ALMA and the JWST --- a general difference observed between these models. No other morphological features can be observed in this object.  

For $z=6$ and 4, we only show the ALMA images for Z6HCW and Z4HCW, respectively (Fig.\,\ref{fig:ALMA2}). The corresponding JWST images of these galaxies are shown in Figures\,\ref{fig:photo_images} and \ref{fig:photo_images2}. Both galaxies show a resolved disk, especially when observed inclined. But only the Z4HCW object hints towards existence of a spiral structure in a disk which spins anti-clockwise.

%%%%%%%%%%%%%%%%%%%%%%%%%%%%%%%%%%%%%%%%%%%%
\subsection{Replacing the redshift by the stellar mass $M_*(z)$}
\label{sec:replace}
%%%%%%%%%%%%%%%%%%%%%%%%%%%%%%%%%%%%%%%%%%%%
    
 To compare evolution of our modeled galaxies with a large observational sample available in the literature, we replaced the explicit dependence on redshift by the dependence on stellar mass, $M_*(z)$, and analyzed the main sequence of star forming galaxies, i.e., SFR[$M_*(z)]$, using stellar mass bins. The main sequence for SF can be approximated by ${\rm log\,SFR} = \alpha {\rm log\,M_* + \beta}$. Some flattening has been claimed to occur in the mass range of $10^{10}-10^{11}\,{\rm M_\odot}$, based on the observed HST Frontier Fields at $1.3\ltorder z < 6$ \citep[e.g.,][]{santi17}, which could be real \citep[e.g.,][]{tom16}. Or it can be related to the difficulty in determining the SFR in deeply dust-obscured high-$z$ massive galaxies \citep[][]{dun17}. 
 
 Figure\,\ref{fig:SFR_Mstar} displays the SFR versus $M_*$. For $z_{\rm f}=6$ galaxies we used the simulated galaxies in the redshift bin of $z\sim 6-5$. For $z_{\rm f}=4$, we used galaxies within $z\sim 4-3$, and for $z_{\rm f}=2$, we used galaxies within $z\sim 2.1-2$.  All galaxies, with the exception of $z_{\rm f}=6$ galaxies with ${\rm log}\,M_*\ltorder 8.7$, lie within one $\sigma$ off the observational black line \citep{santi17}. We confirm that our galaxies trace the relation for the $z_{\rm f}=2$ most massive galaxies with $M_*\gtorder 2\times 10^{10}\,{\rm M_\odot}$.
 
Note that the observational results [the solid line from \citet{santi17}] refer to the main sequence of the galaxy population, while the points represent the small sample of our modeled galaxies which follow the observations. If we trace the same galaxy, marked by the same marker, between each of the sub-figures, the evolutionary track of this galaxy can be recovered. If however, we ignore the slow evolution of the observational main sequence with redshift, and plot the path of any of our galaxies, we find that this path corresponds reasonably well to the observed slope. That means the individual simulated galaxies evolve within $1\sigma-2\sigma$ to the galaxy population.  The close correspondence of these evolutionary tracks to the observed main sequence should not be taken for granted and is not a trivial one.
    
Figure\,\ref{fig:maxvel_Mstar} provides the TF relation, i.e., the maximal rotational velocity, $v_{\rm max}$, of modeled galaxies as a function of $M_*(z)$. The dashed line is the observational average and has a slope of 4.48, obtained from the sample of local field galaxies in the NIR \citep{torr11}. We assumed that the redshift evolution of the $v_{\rm max}-M_*$ slope is negligible \citep[e.g.,][]{port07}, but note that this is disputed at present \citep[e.g.,][]{ubler17}.

We observe that all galaxies, independent of feedback type, increase $v_{\rm max}$ with the stellar mass. The CW galaxies lie well above the VW ones, basically for all stellar masses. For smaller masses, the $v_{\rm max}-M_*$ slope is shallower than 4.48 for $z_{\rm f}=6$ and $z_{\rm f}=4$ galaxies. But for $z_{\rm f}=2$ galaxies,  the slope is very close to 4.48, except for the most massive galaxies $\gtorder 10^{10}\,{\rm M_\odot}$, where the slope steepens.

%%%%%%%%%%%%%%%%%%%%%%%
\section{Discussion and conclusions}
\label{sec:discussion}
%%%%%%%%%%%%%%%%%%%%%%%

This work compares the galaxy evolution constrained in similar mass DM halos towards the final redshifts of $z_{\rm f} = 6$, 4, and 2. Our choice for similar mass halos has been explained in section\,\ref{sec:intro}. {  We expect that higher $z_{\rm f}$ halos grow faster, and this effect propagates down to galactic scales, affecting the stellar feedback and galactic morphology.} We focus on developing morphology of these galaxies, and perform the bulge-disk decomposition based on their stellar surface density, surface photometry, and on stellar kinematics. For brevity, we also provide two additional kinematic stellar spheroid-disk decompositions. All models have been run with high and low overdensity, and with two types of feedback, CW and a stronger VW. We have obtained the synthetic JWST images of these galaxies, post-processed them with a 3-D radiation transfer in the presence of dust, then redshifted, pixelized and convolved with the PSF at all $z_{\rm f}$. These images were compared with the synthetic images obtained by ALMA at $z_{\rm f}$, to test whether underlying morphology can be detected. Furthermore, we have analyzed the stellar migration within the disk-halo system. Finally, we have analyzed the evolution of basic parameters of the modeled galaxies as a function of their stellar mass.  

{  To summarize our main results, we find that 

\begin{itemize} 

\item Although our galaxies evolve in different epochs, their global parameters (e.g., baryonic masses and sizes) remain within a narrow range. On the other hand, their morphology, kinematics and stellar populations differ substantially, yet all of them host sub-kpc stellar bars; 

\item The SFRs appear higher for larger $z_{\rm f}$, and so is their energy and momentum feedback, both for CW and VW objects, all of which follow the main sequence for galaxy SF; 

\item The B/T ratios appear independent of the last merger time for all $z_{\rm f}$, contrary to suggestions in the literature; 

\item The synthetic JWST images include stellar disks at all $z_{\rm f}$, with traces of spiral structure detectable at $z_{\rm f} = 4$ and 2. The CW bars disappear in the PSF-degraded JWST images, but remain visible in the VW models; 

\item Based on the kinematic decomposition, for stellar disks separated from bulges and spheroids, we record a significantly larger rotational support in VW disks compared to CW ones. In both sequences, the disk rotational support increases with decreasing $z_{\rm f}$; 

\item Finally, the ALMA images detect disks at all $z_{\rm f}$, but their spiral structure is only detectable in $z_{\rm f}=2$ galaxies.

\end{itemize}}

Our results have been presented in section\,\ref{sec:results}. Table\,\ref{tab:DMsim} shows $\uplambda$ for our DM only simulations and those with baryons. We find two trends. First, for $z_{\rm f}=6$ halos, the DM only runs exhibit larger $\uplambda$, compared to baryonic runs, while for $z_{\rm f}=4$ and 2, this trend has been reversed. Note that the spin has been calculated only for the particles bound to the DM halos. Second, both the pure DM halos and baryonic DM halos display a higher spin for the large overdensity, $\updelta=3$. The only exclusion is the pure DM halos at $z_{\rm f}=2$, which exhibit no difference between the high and low overdensities.  Moreover, the environment appears to affect the DM halo spin. While the direction of change in the halo spin due to the influx of baryons appears random, the amplitude of this change is substantially larger for larger $\updelta$, by a factor of a few. (The high $\updelta$ objects in our figures lie in the upper row.) This may be interesting and deserves a future attention.  

The largest difference between simulated galaxies comes clearly from the stellar feedback in the form of the galactic wind. The difference in the feedback can be even observed on scales of $\sim 100\,$kpc (in comoving coordinates) in its effect on the cold accretion flow (e.g., Figure\,\ref{fig:inflow}). While the CW feedback creates the pressure bubble around the galaxy, in VW models, the outflow appears to burst through between the incoming filaments of the cold accretion. 

Despite a very different feedback, the baryonic masses of galaxies at the end of the runs are similar, independent of $z_{\rm f}$, and independent of the environment. Therefore, the stellar masses of CW galaxies exceed these of the VW galaxies by a factor of 1.5--2, being compensated by a smaller gas fraction, $f_{\rm gas}$ (Table\,\ref{tab:deco_dens}).  We do not find dependency of $f_{\rm gas}$ on the environment among the CW objects, but do find that $f_{\rm gas}$ is highest for the $z_{\rm f}=6$ CW galaxies, and decreases slightly thereafter.    

\begin{figure}
\center 
	\includegraphics[width=0.4\textwidth]{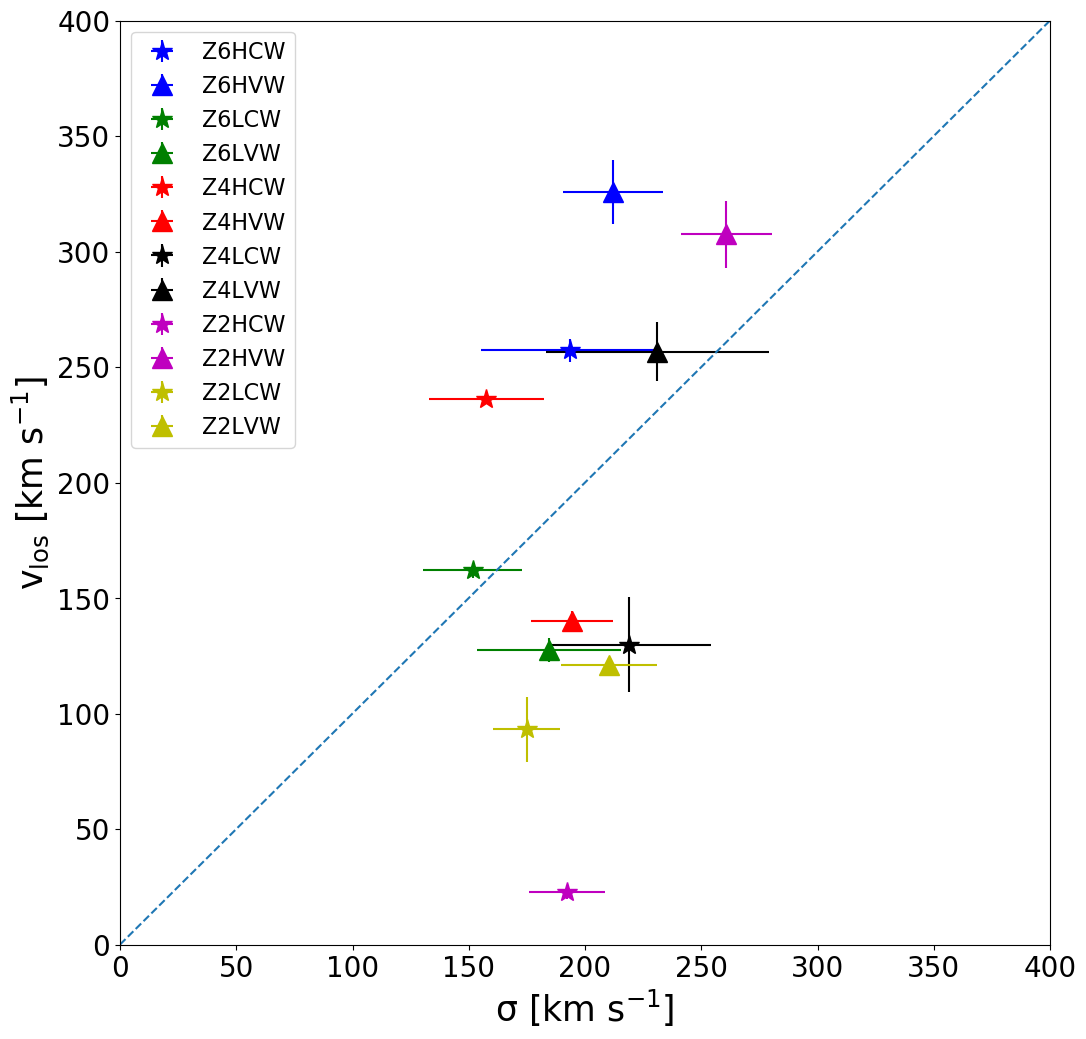}
    \caption{Rotational versus dispersion velocity support for simulated stellar bulges at $z_{\rm f}$, i.e., $v_{\rm los}$ vs $\sigma$. The dashed line corresponds to $v_{\rm los}/\sigma=1$.}
    \label{fig:bulge_support}
    \end{figure} 

  \begin{figure}
\center 
	\includegraphics[width=0.49\textwidth]{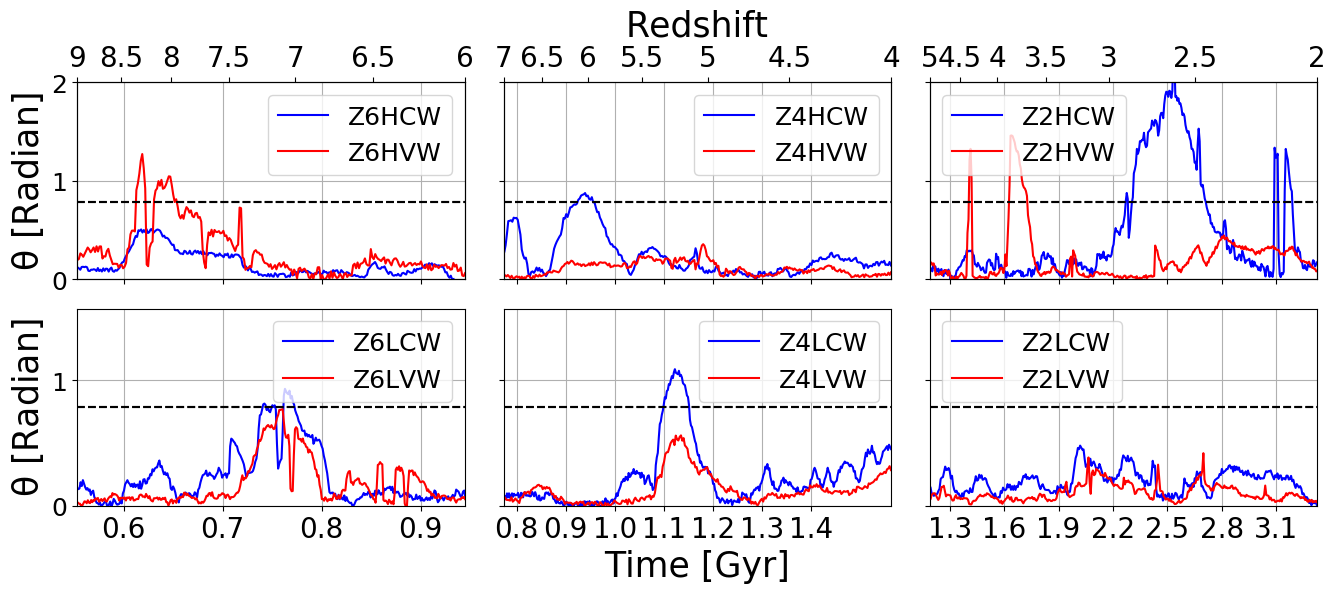}
    \caption{Evolution of the inclination angle between the stellar and gaseous disks. The angle $\theta$ displays the angle between the angular momenta of stars and gas. The dashed line is $\pi/4$.}
    \label{fig:theta}
    \end{figure} 

All modeled galaxies possess a disk component, basically at all times. These disks appear to be much more resilient against destruction even at times of major mergers. This effect can be attributed to disks being very gas-rich and hence able to be restored in short times --- a major difference with the low-$z$ galaxies. The importance of the spheroidal component in our modelled galaxies varies, but the ratio S/T was never found to be greater than unity, and lies within a relatively narrow range of $\sim 0.5-0.8$, as we discuss below. All the disks appear to host sub-kpc stellar bars, more significant in the VW galaxies. Although we address the bar properties in Paper\,II, Figures\,\ref{fig:photo_images} and \ref{fig:photo_images2} allow to assess their detection feasibility at different redshifts using the JWST, and Figures\,\ref{fig:ALMA} and \ref{fig:ALMA2} similarly for the ALMA imaging.

The rotational support for the modeled galaxies appears to be higher for the gaseous component, $v_{\rm los}/\sigma_{\rm z}\sim 1-5$, and varies strongly during the major mergers and close flybys (Fig.\,\ref{fig:vsigma_gas}). No dependence on $z_{\rm f}$ was found. The rotational support for the stellar component is weaker, i.e., $v_{\rm los}/\sigma_{\rm z}\sim 1-3$, and varies less during mergers and flybys. In few cases, this ratio actually decreases to unity at $z_{\rm f}$ (Fig.\,\ref{fig:vsigma_stars}). However, after separating the stellar disk from the bulge and spheroid using kinematic decomposition, the rotational support increases significantly, especially for $z\ltorder 4$ (Table\,\ref{tab:BTcompare}). The temperature of the gas in galaxies is higher in VW galaxies, and we attribute this to a stronger stellar feedback. But overall, the temperature stays flat with redshift. 

While the stellar feedback, CW versus VW, appears to explain the difference in evolutionary patterns of simulated galaxies, we also find that the overdensity provides its contribution as well. Figure\,\ref{fig:theta} exhibits the inclination angles between the stellar and gaseous disks. One can observe that the large angles between the gaseous and stellar disk components spike --- the result of mergers, close flybys and sudden influx of cold accretion material. Galaxies in heavily overdense regions have a larger frequency of these phenomena, which we attribute to the environmental effects.

We have confirmed that our galaxies are "well-behaved" and follow the TF relation, i.e., stellar mass, $M_*$, versus maximal rotation velocity, $v/\sigma$, yet it is evidently not a tight relation. It has been generally recognized that starforming galaxies transit from dispersion-dominated state to rotation-dominated state around $z\sim 1$, due to decrease in mergers and cold accretion inflows which lead to warping and other distortions in the underlying disk \citep[e.g.,][]{kass12}. Our main take from Figure\,\ref{fig:maxvel_Mstar} is that within the mass range analyzed here, $\sim 10^8 - 10^{11}\,{\rm M_\odot}$ the galaxies more massive than $\sim 10^9\,{\rm M_\odot}$ show less scatter on the $M_*-v_{\rm los}$ plane. Furthermore, we observe that our galaxies lie within $1\sigma$ to the SFR main sequence.
 
Figures\,\ref{fig:BD_CWdens} -- \ref{fig:dec_photoVW} display the results of bulge-disk decomposition based on the stellar surface density of modeled galaxies, the surface photometry using the JWST filters, and on the stellar kinematics. These results have been summarized in Tables\,\ref{tab:deco_dens} -- \ref{tab:BTcompare}. The Sersic index for the bulge based on the stellar surface density distinguishes between the CW and VW galaxies --- with a very small overlap, $n_{\rm b} > 0.92$ for the CW galaxies and $n_{\rm b} < 1$ for the VW galaxies, with no dependence on $z_{\rm f}$. Only one galaxy, Z6HCW, exhibits the bulge with $n_{\rm b}\sim 2.27$, the rest have $n_{\rm b} < 2$. 

Using the surface photometry with the JWST filters and convolving with the PSF, we find that $n_{\rm b} < 1$ for the VW objects --- unchanged from the much higher resolution described above. The single exception is the Z6HVW galaxy with $n_{\rm b}\sim 3.7$ --- exhibiting the de\,Vaucouleurs law. For the surface photometry decomposition, we do not find substantial differences between the CW and VW galaxies based on the Sersic index $n_{\rm b}$. Hence, these bulges appear more disklike. In some cases, especially for the VW galaxies, $n_{\rm b}\sim 0.5$, which is symptomatic of galaxies dominated by stellar bars \citep[e.g.,][]{gad09}, as indeed the VW galaxies are. One galaxy, Z4HCW, does not require the bulge component. We do not observe variation with $z_{\rm f}$, nor with the overdensity $\updelta$.

Despite that almost all bulges appear disklike in stellar and photometric decomposition, there is little correspondence between their Sersic indexes, $n_{\rm b}$, which differ sometimes even by a factor of $\sim 5$, with the photometric $n_{\rm b}$ being at the low end --- definitely the consequence of the lower resolution based on the JWST imaging. 

Comparison of all three types of the bulge-to-total mass ratios, B/T (Table\,\ref{tab:BTcompare}), reveals some trends. First, the photometric decomposition results understandably in smaller B/T. The reason for this is that, at all redshifts under consideration here, we are limited to specific bands which are dominated only by a fraction of stars in the rest frames. In this bands, the effect of dust obscuration is substantial and dims the light significantly. Note that the galactic gas is centrally-concentrated, as well as found in the spiral arms, and so the dust obscuration is being far from uniform. Second, both the CW and VW galaxies show very close B/T, differing by less than a factor of 1.5, either in stellar surface density or in surface photometry methods. 

Third, Table\,\ref{tab:BTcompare} shows that for all galaxies, CW and VW, the ratio B/T is always $\gtorder 0.3$ for stellar surface decomposition. This can be compared to the local universe, where observations detect that about 69\% of galaxies with $M_*\gtorder 10^{10}\,{\rm M_\odot}$, have B/T < 0.2 \citep[e.g.,][]{wein09}. However, if we switch to photometric results in Table\,\ref{tab:deco_photo}, we find that irrespective of the feedback type, about 67\% of our galaxies have B/T $< 0.2$. Hence, switching to photometry, which includes the radiation transfer in the presence of dust, redshifting, pixelizing, and convolving woth the PSF, brings the numerical results very close to the observational ones. 
 
We take a more careful look at the galactic bulges obtained using the surface stellar density decomposition by testing their rotational versus dispersion velocity support, $v_{\rm los}/\sigma_{\rm z}$. Are these bulges predominantly disky or classical? Figure\,\ref{fig:bulge_support} shows that the modeled bulges spread much more in $v_{\rm los}$ than in $\sigma_{\rm z}$. They also have a narrow range in their compactness parameter, which corresponds to a narrow range in $\sigma\sim 150-250\,{\rm km\,s^{-1}}$. On the other hand, the maximal rotational velocity spreads on a much wider range, i.e., $v_{\rm los}\sim 20-350\,{\rm km\,s^{-1}}$. 

The bulges in Figure\,\ref{fig:bulge_support} are distributed about symmetrically with respect to the diagonal $v_{\rm los}/\sigma_{\rm z} = 1$. So while the Sersic index describes the bulges as disky, the $v_{\rm los}/\sigma_{\rm z}$ ratio tells a different story --- only half of the bulges appear to be rotationally supported, and the other half, dispersion velocity-supported. It is possible that the decomposition underestimates the values of Sersic indexes for the bulges in some galaxies. We also do not find that the classical bulges, based on their kinematics, have larger B/T than the disky bulges as observed in the local universe \citep[e.g.,][]{drory07}.
 
We have calculated the mass ratio of stellar spheroid-to-total stellar mass, S/T, in galaxies, for all the models at $z_{\rm f}$ (Table\,\ref{tab:BTcompare} and Fig.\,\ref{fig:ST}).   The obtained range, S/T$\sim 0.5-0.8$, is compatible with results from the IllustrisTNG simulations \citep{tacc19}, but the shape of S/T($z$) is flatter compared to their Figure\,2 at $z=2$. Yet they also observe flattening of these curves at $z\ltorder 2$ in the stellar mass range of $10^9-10^{11}\,{\rm M_\odot}$. However, our results indicate some mild increase in S/T towards $z=2$, while the TNG  shows a mild decrease. This difference maybe related to our galaxies residing in the denser environment, $\updelta\sim 3$, and with the associated frequent mergers and flybys, while the TNG displays a statistical average. 
  
Stellar migration reflects the buildup of stellar disk-halo system. We have identified three regions within the DM halos which exhibit clear divisions in the ages of stellar population and the associate density of the stellar component. The two outer regions occupying $\sim 10 - 75\,$kpc in comoving coordinates are characterized by a sharp difference in stellar ages decreasing inwards and stellar densities increasing inwards. These regions consist of stars migrating inwards, i.e., typically obeying $R_{\rm final}\ltorder R_{\rm birth}$. The innermost region inside the $\sim 10$\,kpc, belongs to the growing galaxy, and is characterized by stars migrating outwards, i.e., $R_{\rm final}\gtorder R_{\rm birth}$. The outwards migration is initially nearly nonexistent in the VW galaxies ($z_{\rm f}=6$), but increases sharply at lower redshifts, $z_{\rm f}=4$ and especially in $z_{\rm f}=2$ galaxies. This can be explained by initially a small stellar population in the VW objects, which increases with time and exerts a stronger feedback. This effect is expected to be even stronger at redshifts lower than analyzed here, as both the inward and outward migration appears to  increase with decreasing $z_{\rm f}$, in agreement with \citet{yu20}. 

The growth of stellar mass in galaxies can be described by the PDF of stellar ages (Fig.\,\ref{fig:PDFage}). At larger redshifts one can distinguish between the CW and VW models, but this difference is quickly washed away. As this function reflects the SFRs, one can observe that it increases towards small ages. But the rate of increase depends on $z_{\rm f}$. It is steepest form $z_{\rm f}=6$, less steep for $z_{\rm f}=4$, and becomes flat or even negative for $z_{\rm f}=2$. The maximum is attained at $z\sim 3$.
 
In summary, we used high-resolution zoom-in cosmological simulations to study the emerging morphology and kinematics of galaxies embedded in similar mass DM halos at redshifts 6, 4 and 2, which lie in the high and low overdensity regions. Each galaxy was modeled with two types of stellar feedback, constant velocity wind and a much more energetic variable wind. We find that galaxies at these redshifts are characterized by similar global properties but differ profoundly in the local properties. They have the same baryonic masses, but their stellar mass scales inversely with the stellar feedback. Using three different bulge-disk decomposition methods, namely, the stellar surface density, the surface photometry, and the kinematic properties, we find that the first two methods result in the disk-like bulges, but exhibit a mixed disk-like and classical bulge behavior based on the kinematics. Nevertheless, these bulges form a uniform population based on the Sersic indexes and the B/T mass ratios. The stellar spheroid-to-total masses of these galaxies lie in the range S/T$\sim 0.5-0.8$. 

Performing the photometric bulge-disk decomposition using the JWST mosaic of filters, then downgrading their resolution by convolving with the PSF, we demonstrate the presence of the disk component at all final redshifts. The stellar bars are detected only in the synthetic JWST images of the VW galaxies. The usage of ALMA imaging allows detection of underlying morphology, including the presence of gaseous disks, in all $z_{\rm f}$ galaxies. The bar and the spiral structure are detected only at $z=2$, while only the hint of the spiral structure is detected at $z=4$, and not at $z=6$. Finally, we find that the galaxies lie on the main sequence of SF and follow the Tully-Fisher relation most of their evolution, being separated only by the stellar feedback in the form of galactic winds.   

\section*{Acknowledgements}

We thank Phil Hopkins for providing us with the latest version of the code. We are grateful to Alessandro Lupi for his help with GIZMO, to Peter Behroozi for clarifications about ROCKSTAR, and to Eric F. Jim\'enez-Andrade for his help with the JWST and ALMA imaging. I.S. is grateful for a generous support from the International Joint Research Promotion Program at Osaka University. This work has been partially supported by the Hubble Theory grant HST-AR-14584, and by JSPS KAKENHI grant 16H02163 (to I.S.). The STScI is operated by the AURA, Inc., under NASA contract NAS5-26555. E.R.D. acknowledges support of the Collaborative Research Center 956, subproject C4, funded by the Deutsche Forschungsgemeinschaft (DFG). Simulations have been performed using generous allocation of computing time on the XSEDE machines under the NSF grant TG-AST190016, and by the University of Kentucky Lipscomb Computing Cluster. We are grateful for help by Vikram Gazula at the Center for Computational Studies of the University of Kentucky.

%%%%%%%%%%%%%%%%%%%%%%%%%%%%%%%%%%%%%%%%%%%%%%%%%%
\section*{Data Availability}

The data used for this paper can be available upon request. 
 
%%%%%%%%%%%%%%%%%%%% REFERENCES %%%%%%%%%%%%%%%%%%

% The best way to enter references is to use BibTeX:

% Alternatively you could enter them by hand, like this:
% This method is tedious and prone to error if you have lots of references
%\begin{thebibliography}{99}
%\bibitem[\protect\citeauthoryear{Author}{2012}]{Author2012}
%Author A.~N., 2013, Journal of Improbable Astronomy, 1, 1
%\bibitem[\protect\citeauthoryear{Others}{2013}]{Others2013}
%Others S., 2012, Journal of Interesting Stuff, 17, 198
%\end{thebibliography}

%%%%%%%%%%%%%%%%%%%%%%%%%%%%%%%%%%%%%%%%%%%%%%%%%%

%%%%%%%%%%%%%%%%% APPENDICES %%%%%%%%%%%%%%%%%%%%%
\newpage

\appendix

\section{The bulge-disk decomposition}

\subsection{The bulge-disk decomposition based on the stellar surface density}

We obtained the bulge and disk components by fitting the 1-D face-on stellar surface density profile with a model combined the Sersic function and an exponential for the outer disk,
\begin{equation}
\Sigma(r)= \Sigma_{\rm e} e^{-b_{\rm n}[(R/R_{\rm e})^{1/n}-1]} + \Sigma_0 e^{-R/R_{\rm d}}
\end{equation}
where, $\Sigma_{\rm e}$ and $R_{\rm e}$ are the effective surface density and effective radius of the bulge component.  $n$ is the Sersic index of the bulge component. $\Sigma_0$ and $R_{\rm d}$  are the central surface density and scalelength of the exponential disk component. Both components are fitted simultaneously to satisfy the Maximum likelihood estimate for the 1-D face-on projection surface density profile. We added an additional constrain that $r_{\rm d} > r_{\rm e}$ to make sure that the disk component is always outside the bulge component.

\begin{figure}
%\center 
\includegraphics[width=0.43\textwidth]{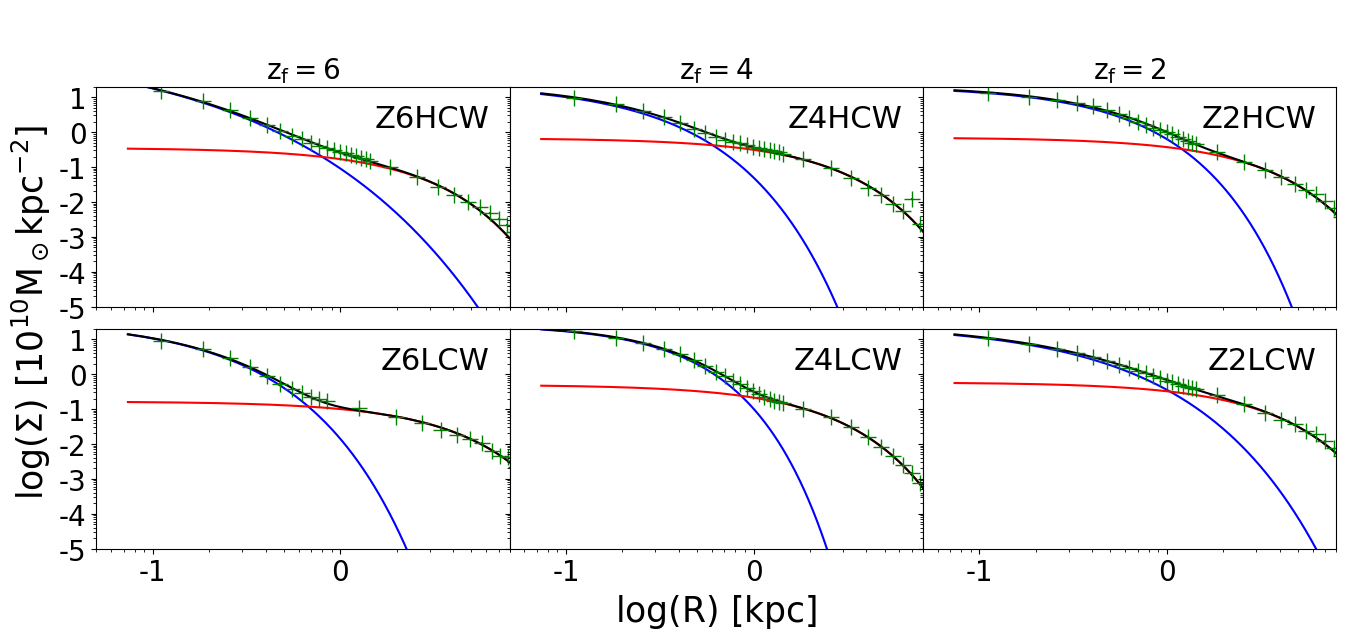}
    \caption{Bulge-disk decomposition of CW models based on the stellar surface density. The sizes are in comoving coordinates.}
    \label{fig:BD_CWdens}
    \end{figure}

\begin{figure}
%\center 
\includegraphics[width=0.43\textwidth]{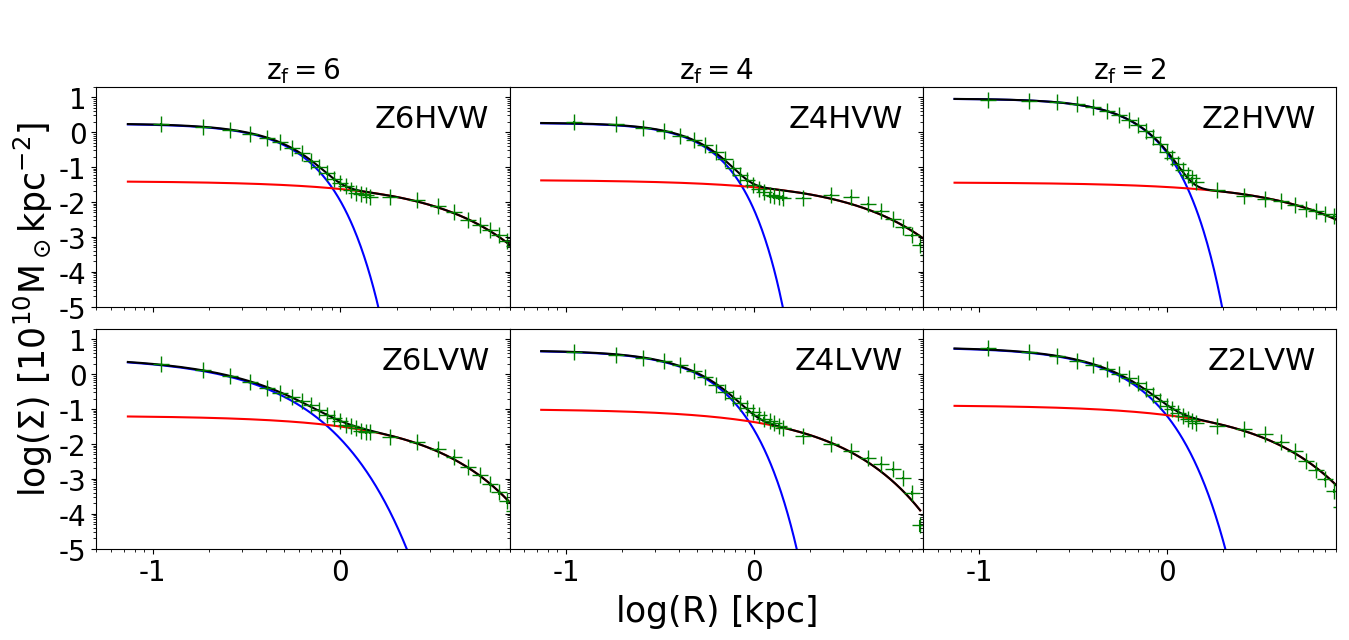}
    \caption{Bulge-disk decomposition of VW models based on the stellar surface density. The sizes are in comoving coordinates.}
    \label{fig:BD_VWdens}
    \end{figure} 

\subsection{Bulge-disk decomposition based on the surface photometry}
 
Similar to the stellar surface density decomposition. For the surface photometry, we have replaced all the mass surface densities with photometric surface brightness,
\begin{equation}
I(r)= I_{\rm e} e^{-b_{\rm n}[(R/R_{\rm e})^{1/n}-1]} + I_0 e^{-R/R_{\rm d}}, 
\end{equation}
where, $I_{\rm e}$ and $R_{\rm e}$ are effective surface brightness and effective radius of the bulge component. $n$ is the Sersic index of the bulge component. $I_0$ and $R_{\rm d}$  are central surface brightness and scale length of the exponential disk component. 

\begin{figure}
\center 
\includegraphics[width=0.43\textwidth]{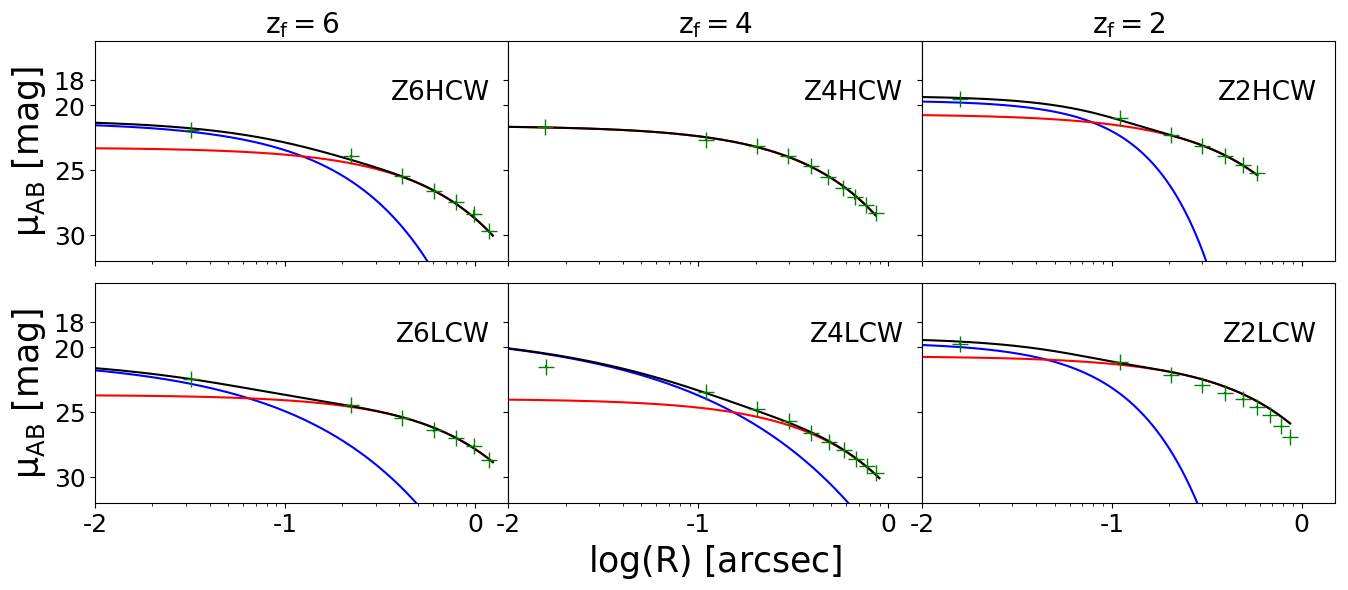}
    \caption{Bulge-disk decomposition for the CW models based on their photometric images using JWST filter F356W for $z_{\rm f}=6$, and filter F200W for $z_{\rm f}=4$ and 2.
    }
    \label{fig:dec_photoCW}
    \end{figure}

\begin{figure}
\center 
\includegraphics[width=0.43\textwidth]{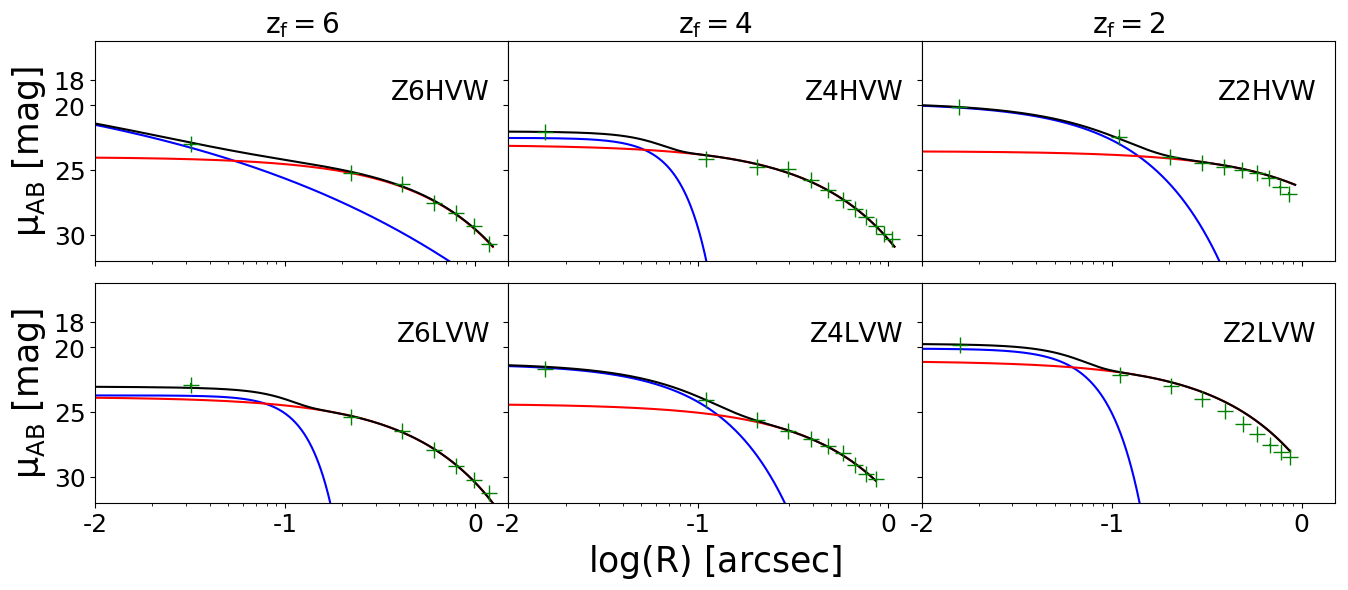}
    \caption{Bulge-disk decomposition for the VW models (as in Figure\,\ref{fig:dec_photoCW})}
    \label{fig:dec_photoVW}
    \end{figure}

%%%%%%%%%%%%%%%%%%%%%%%%%%%%%%%%%%%%%%%%%%%%%%%%%%

% Don't change these lines
\bsp	% typesetting comment
\label{lastpage}

\begin{thebibliography}{99}
%\bibliographystyle{mnras}
%\bibliography{example} % if your bibtex file is called example.bib

%\bibliographystyle{mnras}
%\begin{thebibliography}{99}

\bibitem[Abadi et al.(2003a)]{abadi03a}
 Abadi, M.G., Navarro, J.F., Steinmetz, M., et al. 2003a, \apj, 591, 499 
 
\bibitem[Abadi et al.(2003b)]{abadi03b}
 Abadi, M.G., Navarro, J.F., Steinmetz, M., et al. 2003b, \apj, 597, 21 

\bibitem[Baes et al.(2003)]{baes03}
 Baes, M., Davies, J.I., Dejonghe, H., et al. 2003, \mnras, 343, 1081

\bibitem[Behroozi et al.(2012)]{behr12}
 Behroozi P.S., Wechsler, R.H., Wu, H.Y., et al. 2012, \apj, 763, 18

\bibitem[Behroozi et al.(2013)]{behr13}
 Behroozi, P.S., Wechsler, R.H., Wu, H.-Y. 2013, \apj, 762, 109

\bibitem[Bi et al.(2021)]{bi21}
 Bi, D., Shlosman, I., Romano-Diaz, E. 2021, submitted (Paper\,II)

\bibitem[Binney \& Tremaine(2008)]{binn08} 
Binney, J., \& Tremaine, S. 2008, Galactic Dynamics, Princeton Univ. Press  

\bibitem[Bouwens et al.(2015)]{bouw15}
Bouwens, R.J., et al. 2015, \apj, 803, 34

\bibitem[Bouwens et al.(2017)]{bouw17}
Bouwens, R.J., Oesch, P.A., Illingworth, G.D., Ellis R.S., Stefanon, M., 2017, \apj, 843, 129

\bibitem[Brinchmann \& Ellis(2000)]{brinch00} 
Brinchmann, J., \& Ellis, R.S. 2000, \apjl, 536, L77 

\bibitem[Brooks et al.(2009)]{brooks09} 
Brooks, A.M., Governato, F., Quinn, T., et al. 2009, \apj 694, 396

\bibitem[Bruzual \& Charlot(1993)]{bruz93} 
Bruzual, A.G., \& Charlot, S. 1993, \apj, 405, 538 

\bibitem[Camps et al.(2016)]{camps16} 
Camps, P., Trayford, J.W., Baes, M., et al. 2016, \mnras, 462, 1957

\bibitem[Cassata et al.(2011)]{cass11}
Cassata, P., Le Fevre, O., Garilli, B., et al. 2011, A\&A, 525, 143

\bibitem[Chabrier(2003)]{chab03}
Chabrier, G. 2003, \apjl, 586, L133

\bibitem[Childress et al.(2014)]{child14} Childress, M.J., Wolf, C., Zahid, H.J., 2014, \mnras, 445, 1898 

\bibitem[{{Collier} {et al.}(2018)}]{coll18}{Collier}, A., {Shlosman}, I., {Heller}, C. 2018, \mnras, 476, 1331

\bibitem[Crain et al.(2015)]{crain15}
Crain, R.A., et al. 2015, \mnras, 450, 1937

\bibitem[Croft et al.(2009)]{croft09}
Croft, R.A.C., Di Matteo, T., Springel, V., et al. 2009, \mnras, 400, 43

\bibitem[Dekel \& Birnboim(2006)]{dekel06}  
Dekel, A., Birnboim, Y. 2006, \mnras, 368, 2 

\bibitem[Delgado-Serrano et al.(2010)]{delga10}
 Delgado-Serrano, R., Hammer, F., Yang, Y.B., Puech, M., Flores, H., \& Rodrigues, M. 2010, 
 A\&A, 509, 78 

\bibitem[de Vaucouleurs(1948)]{vau48}
de Vaucouleurs, G. 1948, Annales d'Astrophysique, 11, 247
 
\bibitem[Di Matteo et al.(2008)]{dimatt08}
Di Matteo T., Colberg J., Springel V., et al. 2008,apj, 676, 33
 
\bibitem[Drory \& Fisher(2007)]{drory07}  
Drory, N., Fisher, D.B. 2007, \apj, 664, 640 

\bibitem[Dunlop et al.(2017)]{dun17}
Dunlop, J.S., McLure, R.J., Biggs, A.D., et al. 2017, \mnras, 466, 861

\bibitem[Eisenstein \& Hut(1998)]{eise98}  
Eisenstein, D.J., \& Hut, P. 1998, \apj, 498, 137

\bibitem[El-Badry et al.(2018)]{badry18} 
 El-Badry, K., Quataert, E., Wetzel, A., et al. 2018, \mnras, 473, 1930 

\bibitem[Faucher-Giguere et al.(2009)]{Faucher09} 
 Faucher-Giguere, C.A., Lidz, A., Zaldarriaga, M., Hernquist, L. 2009, \apj, 703, 1416  

\bibitem[F\"orster-Schreiber et al.(2006)]{forst06} 
F\"orster-Schreiber, N.M., et al., 2006, \apj, 645, 1062

\bibitem[Gadotti(2009)]{gad09}
Gadotti, D.A. 2009, in {\it Chaos in Astronomy}, G. Contopoulos, P.A. Patsis (eds.), 
    Springer-Verlag: Berlin, p.159

\bibitem[Genel et al.(2015)]{genel15} 
Genel, S., Fall, S.M., Hernquist, L., et al., 2015, \apj, 804, L40

\bibitem[Genzel et al.(2008)]{genz08} 
Genzel, R., et al., 2008, \apj, 687, 59

\bibitem[Governato et al.(2007)]{gover07} 
 Governato, F., Willman, B., Mayer, L., et al. 2007, \mnras, 374, 1479 
 
\bibitem[Guedes et al.(2011)]{gue11} Guedes, J., Callegari, S., Madau, P., Mayer, L.
  2011, \apj, 742, 76 
 
\bibitem[{{Hahn} \& {Abel}(2011)}]{hahn11}
{Hahn}, O., {Abel}, T. 2011, \mnras, 415, 2101 

\bibitem[Harikane et al.(2018)]{hari18} Harikane Y., et al. 2018, \pasj, 70, S11

\bibitem[{{Hopkins}(2017)}]{hopk17}{Hopkins}, P.F. 2017, eprint 1712.01294 

\bibitem[{{Hopkins et al.}(2018)}]{hopk18}{Hopkins}, P.F. et al. 2018, \mnras, 480, 800 

\bibitem[Ishigaki et al.(2018)]{ishi18}
Ishigaki, M., Kawamata, R., Ouchi, M., Oguri, M., Shimasaku, K.,Ono, Y. 2018, \apj, 854, 73

\bibitem[Jagvaral et al.(2021)]{jag21}
Jagvaral, Y., Campbell, D., Mandelbaum, R., et al. 2021, \mnras, subm., ArXiv:2105.02237:

\bibitem[Kassin et al.(2012)]{kass12}
 Kassin, S.A., Weiner, B.J., Faber, S.M., et al. 2012, \apj, 758, 106

\bibitem[Kawamata et al.(2016)]{kawa16}
Kawamata R., Oguri M., Ishigaki M., Shimasaku K., Ouchi M.,2016, \apj, 819, 114

\bibitem[Keres et al.(2005)]{keres05}
 Keres, D., Katz, N., Weinberg, D.H. \& Dave, R. 2005, \mnras, 363, 2 

\bibitem[Keres et al.(2009)]{keres09}
Keres, D., Katz, N., Fardal, M., et al. 2009, \mnras, 395, 160

\bibitem[Konno et al.(2014)]{konno14}
Konno, A., Ouchi, M., Ono, Y., K., et al. 2014, \apj 797, 16

\bibitem[{{Kormendy} \& {Kennicutt}(2004)}]{kor04}
{Kormendy}, J., {Kennicutt}, R.C. 2004, \araa, 42, 603 

\bibitem[Landau \& Lifshitz(1960)]{lan60}
Landau, L.D., Lifshitz, E.M. 1960, Mechanics, Vol.1, Pergamon Press

%\bibitem[Leitherer et al.(1999)]{leit99}
%Leitherer, C.,  Schaerer, D., Goldader, J.D.,  et al. 1999, \apjs, 123, 3

\bibitem[Lotz et al.(2017)]{lotz17}
Lotz, J.M., et al., 2017, \apj, 837, 97

\bibitem[Madau \& Dickinson(2014)]{madau14} 
Madau, P., Dickinson, M. 2014, \araa, Annu. Rev., 52, 415

\bibitem[Marinacci et al.(2014)]{mari14}
Marinacci, F., Pakmor, R., Springel, V. 2014, \mnras, 437, 1750

\bibitem[Martin et al. (2015)]{mart15}
Martin, D.C., Matuszewski, M., Morrissey, P., et al. 2015, \nat, 524, 192

\bibitem[Muratov et al. (2015)]{mura15}
Muratov, A.L., Keres, D., Faucher-Giguere, C.-A., et al. 2015, \\mnras, 454, 2713

\bibitem[{Navarro {et~al.}(1996)}]{nfw96}
{Navarro}, J.F., Frenk, C.S., White, S.D.M. 1996, \apj, 462, 563 (NFW)

\bibitem[{Nelson {et~al.}(2018)}]{nel18}
{Nelson}, D., et al. 2018, \mnras, 475, 624

\bibitem[Navarro \& White(1994)]{nava94} 
Navarro, J.F., White, S.D.M. 1994, \mnras, 267, 401

\bibitem[{Nelson {et~al.}(2019)}]{nel19}
{Nelson}, D., et al. 2019, \mnras, 490, 3234

\bibitem[Oppenheimer \& Dave(2006)]{oppi06} 
 Oppenheimer, B.D., \& Dave, R. 2006, \mnras, 373, 1265

\bibitem[Oesch et al.(2016)]{oesch16} 
Oesch, P., et al. 2016, \apj, 819, 129 

\bibitem[Ono et al.(2018)]{ono18} Ono, Y., et al. 2018, \pasj, 70, S10

\bibitem[Ott et al.(2012)]{ott12} 
Ott, J., et al. 2012, \aj, 144, 123 

\bibitem[Pallottini et al.(2017)]{pallo17} 
 Pallottini, A., ferrara, A., Gallerani, S., et al. 2017, \mnras, 465, 2540 

\bibitem[Pillepich et al.(2019)]{pille19} 
 Pillepich, A., Nelson, D., Springel, V., et al. 2019, \mnras, 490, 3196 

\bibitem[Plank Collaboration et al.(2016)]{planck16} 
 Planck Collaboration et al. 2016, A\&A, 594, A13 

\bibitem[Portinari \& Sommer-Larsen(2007)]{port07}
Portinari, L., Sommer-Larsen, J. 2007, \mnras, 375, 913

\bibitem[Posti \& Helmi(2019)]{posti19}
Posti, L., Helmi, A. 2019, A\&A, 621, 56

\bibitem[Rauch et al.(2011)]{rau11} 
 Rauch, M., Becker,G.D., Haehnelt, M.G., et al. 2011, \mnras, 418, 1115 

\bibitem[Romano-D{\'{\i}}az et al.(2009)]{roma09} 
 Romano-D{\'{\i}}az, E., et al. 2009, \apj, 702, 1250 

\bibitem[Romano-D{\'{\i}}az et al.(2011)]{roma11} 
 Romano-D{\'{\i}}az, E., Choi, J.-H., Shlosman, I., \& Trenti, M. 2011, \apjl, 738, L19 

\bibitem[Romano-D{\'{\i}}az et al.(2014)]{roma14}
 Romano-D{\'{\i}}az, E., Shlosman, I.,  Choi, J.-H., \& Sadoun, R. 2014, \apjl, 790, L32

\bibitem[Rosas-Guevara et al.(2020)]{rosa20}
 Rosas-Guevara, E., Bonoli, S., Dotti, M., et al. 2020, \mnras, 491, 2547

\bibitem[Sancisi et al.(2008)]{sanci08}
Sancisi, R., Fraternali, F., Oosterloo, T., et al. 2008, A\&A Rev., 15, 189 

\bibitem[Santini et al.(2017)]{santi17}
 Santini, P., Fontana, A., Castellano, M., et al. 2017, \apj, 847, 76

\bibitem[Scannapieco et al.(2009)]{scanna09}
Scannapieco, C., White, S.D.M., Volker, S., Tissera, P.B. 2009, \mnras, 396, 708

\bibitem[Schaye et al.(2015)]{schaye15}
Schaye, J., et al., 2015, \mnras, 446, 521

\bibitem[Shibuya et al.(2021)]{shibu21}
 Shibuya, T., Miura, N., Iwadate, K. 2021, arViv eprints 2106.03728

\bibitem[{{Shlosman}(2013)}]{shlo13}{Shlosman}, I. 2013, in Secular Evolution of Galaxies, 
CUP, (eds.) J.Falcon-Barroso \& J.H.Knapen, p.555 (arXiv:1212.1463)

\bibitem[Simons et al.(2015)]{simo15}
 Simons, R.S., Kassin, S.A., Weiner, B.J., et al. 2015, \apj, 452, 986

\bibitem[Simons et al.(2017)]{simo17}
Simons, R.C., et al., 2017, \apj, 843, 46

\bibitem[Springel \& Hernquist(2003)]{spri03} Springel, V., \& Hernquist, L. 
 2003, \mnras, 339, 289 

\bibitem[Springel(2005)]{spri05} Springel, V. 2005, \mnras, 364, 1105

\bibitem[Springel et al.(2008)]{spri08} 
Springel, V., Wang, J., Vogelsberger, M., et al. 2008, \mnras, 391, 1685 

\bibitem[Simons et al.(2017)]{simo17}
Simons, R.C., et al., 2017, \apj, 843, 46

\bibitem[Tacchella et al.(2019)]{tacc19}
Tacchella, S., Diemer, B., Hernquist, L., et al. 2019, \mnras, 487, 5416

\bibitem[Tomczak et al.(2016)]{tom16}
Tomczak, A.R., Quadri, R.F., Tran, K.-V.H., et al. 2016, \apj, 817, 118

\bibitem[Torres-Flores et al.(2011)]{torr11}
Torres-Flores, S., Epinat, B., Amram, P., et al. 2011, \mnras, 416, 1936

\bibitem[Tully \& Fisher(1977)]{tully77}
 Tully, R.B., Fisher, J.R. 1977, A\&A, 54, 661 

\bibitem[\"Ubler et al.(2017)]{ubler17}
\"Ubler, H., F\"orster-Schreiber, N.M., Genzel, R., et al. 2017, \apj, 842, 121

\bibitem[van de Voort et al.(2012)]{voort12}
van de Voort, F., Schaye, J., Altay, G., et al. 2012, \mnras, 421, 2809

\bibitem[Vernet et al.(2017)]{ver17}
Vernet, J., Lehnert, M.D., De Breuck, C., et al. 2017, A\&A, 602, L6

\bibitem[{{Vogelsberger} {et~al.}(2014)}]{vogel14}
{Vogelsberger}, M., et al. 2014, \mnras, 444, 1518  

\bibitem[Vogelsberger et al.(2018a)]{vogel18a}
Vogelsberger, M. et al. 2018a, \mnras, 444, 1518

\bibitem[Vogelsberger et al.(2018b)]{vogel18b}
Vogelsberger, M. et al. 2018b, Nature, 509, 177

\bibitem[Walter et al.(2008)]{walt08}
Walter F., Brinks E., de Blok W.J.G., et al. 2008, \aj, 136, 2563

\bibitem[Weinzirl et al.(2009)]{wein09}
 Weinzirl, T., Jogee, S., Khochfar, S., Burkert, A., \& Kormendy, J. 2009, \apj, 696, 411

\bibitem[White \& Rees(1978)]{white78}
 White, S.M.D., Rees, M.J. 1978, \mnras, 183, 341

\bibitem[Yu et al.(2020)]{yu20} 
 Yu, S., et al. 2020, \mnras, 494, 1539 


\end{thebibliography}
\end{document}